# CONCEPTUAL AND MATHEMATICAL MODELING OF INSECT-BORNE PLANT DISEASES: THEORY AND APPLICATION TO FLAVESCENCE DORÉE IN GRAPEVINE

## RESEARCH REPORT R947

**FEDERICO MAGGI**

The University of Sydney, School of Civil Engineering, 2006 Sydney, NSW, Australia.
ISI Foundation, Viale S. Severo 65, 10133 Torino, Villa Gualino, Italy.

**DOMENICO BOSCO**

Università di Torino, DISAFA., Entomologia e Zoologia Applicate all'Ambiente, Grugliasco, TO, Italy.

**CRISTINA MARZACHÌ**

Istituto di Virologia Vegetale, CNR, Torino, Italy

February, 2014

**Copyright Notice**

School of Civil Engineering, Research Report R947

Conceptual and mathematical modeling of insect-borne plant diseases: theory and application to Flavescence dorée in grapevine

Federico Maggi
Domenico Bosco
Cristina Marzachi'

February 2014

ISSN 1833-2781

Published by:
School of Civil Engineering
The University of Sydney
Sydney NSW 2006
Australia

This report and other Research Reports published by the School of Civil Engineering are available at http://sydney.edu.au/civil

## ABSTRACT

Insect-borne plant diseases recur commonly in wild plants and in agricultural crops, and are responsible for severe losses in terms of produce yield and monetary return. Mathematical models of insect-borne plant diseases are therefore an essential tool to help predicting the progression of an epidemic disease and aid in decision making when control strategies are to be implemented in the field. While retaining a generalized applicability of the proposed model to plant epidemics vectored by insects, we specifically investigated the epidemics of Flavescence dorée phytoplasma (FD) in grapevine plant *Vitis vinifera* specifically transmitted by the leafhopper *Scaphoideus titanus*. The epidemiological model accounted for life-cycle stage of *S. titanus*, FD pathogen cycle within *S. titanus* and *V. vinifera*, vineyard setting, and agronomic practices. The model was comprehensively tested against biological *S. titanus* life cycle and FD epidemics data collected in various research sites in Piemonte, Italy, over multiple years. The work presented here represents a unique suite of governing equations tested on existing independent data and sets the basis for further modelling advances and possible applications to investigate effectiveness of real-case epidemics control strategies and scenarios.

# TABLE OF CONTENTS

# I. INTRODUCTION

## I.1. BACKGROUND

Flavescence Dorée (FD) is the most important and destructive phytoplasma disease of *Vitis vinifera* grapevines, and is characterized by severe symptoms such as leaf roll with yellowing (in white varieties) and reddening (in red varieties), withering of bunches, lack of cane lignification, shortening of internodes with general stunting, and reduced grape production. FD is transmitted by the grapevine specialist leafhopper *Scaphoideus titanus* Ball and can spread so rapidly in the absence of control measures to infect the totality of vines within a vineyard in only few years.

The Regione Piemonte has been undergoing an important FD outbreak since the late nineties, which is threatening the grape production with important consequences on several agricultural, economical, and social implications. FD epidemics depend on the characteristics and complex relationships of the coupled phytoplasma-insect-plant system (Biere and Bennet, 2013). These complex interactions are reflected in highly nonlinear epidemiological processes most of which are still not clearly known. The major reason for a lack in the understanding are the time scales involved in the progression of the epidemics, which typically occur over multiple years. Furthermore, a number of biotic (e.g., grapevine cultivar type, vector mobility and life cycle) and abiotic factors (e.g., agricultural practices) affects FD epidemics progression and results in different epidemics patterns. These characteristics have hampered a systematic and detailed investigation of the cycle and progression of FD epidemics across the Regione Piemonte. The fundamental issue is that FD epidemics is still active in the region even if mandatory insecticide treatments have been applied since 2001 (Ministry Resolution, DM 32442, 31st May, 2000).

The state-of-the-art understanding of FD dynamics mostly relies on experimental observations, whereas mathematical modeling has not been attempted so far for this phytoplasma. Together with epidemics monitoring, mathematical tools are instrumental to predict the epidemics progression, assess the effectiveness of remediation strategies, and interpret scenario-related outbreak risks (Shtienberg, 2000). Models of phytovirus epidemics have been proposed in the past twenty years but these have not been generalized to phytoplasmas and most of them are not suitable for FD epidemics in grapevine. In addition, and regardless of the specific pathogenic microorganism, the current models of plant diseases do not include the simultaneous life cycle of plant host and insect vector as yet (Jeger et al., 2004; Madden et al., 2000; Nakazawa et al., 2012), neither have they been tested on real epidemiological data (Madden and van den Bosch, 2002) with the exception of the stochastic model in Ferriss and Berger (1993). These aspects leave a serious gap in methods that can be used to address the epidemics predictability, outbreak risk evaluation, and scenario analysis in a systematic way.

## I.2. AIMS

The aim of this work is to propose a better understanding of the FD epidemics currently occurring in the Regione Piemonte. To this end, we bridged experimental recordings and numerical modeling to develop and partly test a mathematical framework of FD epidemics to aid in policy and decision making. Specifically, we report on development and integration of the governing equations describing *S. titanus* dynamics (*e.g.*, life stages, longevity, prolificacy, etc.), the interactions between FD phytoplasma, *S. titanus* and *V. vinifera* (*e.g.*, latency, insect mobility and transmission efficiency), and the vector and host response to FD phytoplasma (*e.g.*, reduced fitness, mortality, recovery).

# II. MATHEMATICAL MODELING

## II.1 FD DISEASE PATHWAYS AND MODEL STATE VARIABLES

The FD phytoplasma is a wall-less bacterium belonging to the 16S ribosomal group V, subgroups C and D, and the establishment of the "*Candidatus* Phytoplasma vitis" species has been suggested (IRPCM, 2004).

FD phytoplasma develops between the vector leafhopper *S. titanus* and *V. vinifera* plant along the pathways depicted in Figure 1. *S. titanus* has a typical monovoltine life cycle that includes eggs, I-II instar nymphs without wing pads, III-V instar nymphs with wing pads (poorly mobile compared to adults), and winged and flying adults (mobile vectors). Each stage occurs over a time window defined by time of initial development, duration, and sink rates (e.g., Bosio and Rossi, 2001). Healthy *S. titanus* leafhoppers acquire FD phytoplasma while feeding on infective plants but are not able to transmit FD to grapevines until after a latency period $L_I$ (e.g., Schvester et al., 1963; Caudwell and Larrue, 1986; Caudwell et al., 1987), after which they remain infective for life. FD affects *S. titanus* fitness by reducing its longevity and fecundity (Bressan et al., 2005). Healthy grapevines may become infected after infective *S. titanus* feeds on them injecting FD phytoplasmas into the phloem with saliva. The latent period $L_P$ in plants normally results in systemic infection with symptoms in the year after FD inoculation. Affected grapevine plants can be either persistently infected (and eventually die) or recover (Morone et al., 2007; Musetti et al., 2007; Galetto et al., 2009). Thus, the FD cycle continues anew when infective plants are a suitable source of infection for healthy insects (Figure 1).

*S. titanus* dynamics can be described by the following state variables (Maggi et al., 2013): number of eggs $E_n$ hatching in year $n$, number $N_1$ of I-II instar nymphs, number $N_2$ of III-V instar nymphs, number of adults $I$, and the number of laid eggs $E_{n+1}$ that will hatch in year $n + 1$ (Figure 1). The number of infected adults $I_i$ (including latently-infected and infective) and the number of infective adults $I_t$ (only those that can transmit FD) are the state variables describing FD infection in *S. titanus*. The state variables describing FD epidemics in *V. vinifera* are the total number of plants $P$, the number $P_i$ of infected plants (including latently-infected and infective), and the number $P_t$ of infective plants (from which vectors can acquire FD).

The relations between state variables $I \geq I_i \geq I_t$ in *S. titanus* and $P \geq P_i \geq P_t$ in *V. vinifera* hold as per the sets depicted in Figure 1, and allow us to write the governing equations of FD epidemics in a convenient mathematical form.

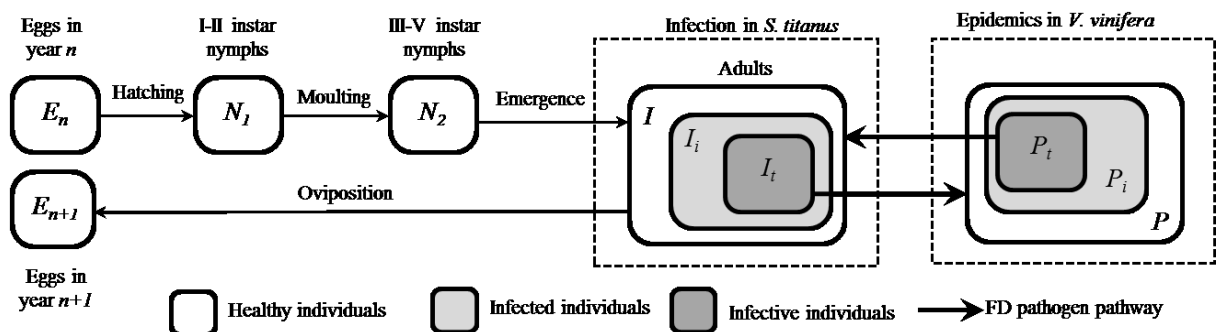

**Figure 1**. FD phytoplasma pathway across *S. titanus* leafhopper and *V. vinifera* plant. Insect groups defined by $E_n$, $N_1$, $N_2$, $I$, $I_i$, $I_t$, and $E_{n+1}$, and the plant groups $P$, $P_i$ and $P_t$ correspond to the insect and plant state variables used to model FD epidemics.

## II.2 GOVERNING EQUATIONS OF *S. TITANUS* DYNAMICS

*S. titanus* has a typical univoltine life cycle and overwinters as egg. Its life cycle develops through seven stages from eggs, to I-II instar nymphs (without wing pads), III-V instar nymphs (with wing pads), and adult (winged). The mobility of nymphs is very low compared to that of adults, which represent the mobile flying vectors. Each stage occurs over a time window defined by time of initial development and duration, the latter being better described by the turnover time (i.e., a duration) inversely proportional to the rates at which individuals in one biological stage step into the next (e.g., Bosio and Rossi 2001).

To describe *S. titanus* dynamics, the state variables introduced in Section II.1 were use, where I-II and III-V instar nymphs were aggregated in $N_1$ and $N_2$, respectively, as the differences are minor and not relevant toward FD spread. FD acquisition from *V. vinifera* normally occurs at the stage of $N_2$ and $I$ (Figure 1), since newly hatched nymphs may not feed in the phloem because of their short stylets, and are unlikely to acquire phytoplasmas (Carle and Moutous 1965). However, only *S. titanus* adults represent a source of infection for *V. vinifera* because FD requires a latency period of about 30 days in the vector before it can be transmitted to plants, and because adults are much more mobile than nymphs.

The five-stage size-dependent *S. titanus* life-cycle model can be described by the following piecewise continuous-in-time ordinary and integro-differential equations for the yearly time scale as (Maggi et al., 2013)

$$\frac{dE_n(t)}{dt} = -H(t - t_E) \times \varepsilon E_n(t) \left[ 1 - \frac{E_n(t)}{K(t)} \right] - p E_n(t) \ , \tag{1}$$

$$\frac{dN_1(t)}{dt} = H(t - t_E) \times \varepsilon E_n(t) \left[ 1 - \frac{E_n(t)}{K(t)} \right] - p N_1(t) - H(t - t_{N_1}) \eta_1 N_1(t) \ , \tag{2}$$

$$\frac{dN_2(t)}{dt} = H(t - t_{N_1}) \times \eta_1 N_1(t) - p N_2(t) - H(t - t_{N_2}) \eta_2 N_2(t) \ , \tag{3}$$

$$\frac{dI(t)}{dt} = H(t - t_{N_2}) \times \eta_2 N_2(t) \ - \delta I(t) - p I(t) - y \rho_{YST} S I(t) \ , \tag{4}$$

$$\frac{dE_{n+1}(t)}{dt} = H(t - t_L) \times \rho [1 - f_{MF}(t)] I(t) - p E_{n+1}(t) \ , \tag{5}$$

where $H(X)$ in Eq. (1) to Eq. (5) is the Heaviside step function (i.e., $H(X) = 1$ for $X \geq 0$, $H(X) = 0$ otherwise) which defines the timing of $E_n$ hatching, $N_1$ and $N_2$ moulting, $I$ emergence, and egg deposition. The rate constant $p$ represents abiotic and biotic sinks such as predation, parasitism, environmentally-induced mortality, *etc.*, and was identically applied to all *S. titanus* stages, $\varepsilon$ is the egg hatching rate, $\eta_1$ and $\eta_2$ are the moulting and emergence rates from $N_1$ to $N_2$ nymphs and from $N_2$ nymph to adult $I$, respectively, $\delta$ is the adult mortality rate, $\rho$ is the female egg deposition rate, and $f_{MF}$ is the male-to-female sex ratio. The times $t_E$, $t_{N_1}$, $t_{N_2}$ and $t_L$ define the hatching, moulting, emergence, and oviposition timing, respectively, and satisfy $t_E < t_{N_1} < t_{N_2} < t_L$ as from observations.

In Eq. (1), the number of eggs decreases over time due to hatching according to logistic kinetics (e.g., Murray, 2002) and by a first-order kinetics sink due to abiotic and biotic factors (e.g., Ngwa 2006). The time-dependent variable $K(t) = E_{n0} - \int_0^t p E_n(t^*) dt^*$ is the system carrying capacity representing the net number of individuals that contribute egg hatching to nymph after abiotic and biotic sinks out of an initial number $E_{n0}$ in year $n$. The logistic equations in Eq. (1) and Eq. (2) are solved for $t \geq t_E$ but with the prescription that hatching at time $t = t_E$ occurs to only 1 egg, and results in the increment of 1 nymph $N_1$. This is mathematically written as $E_n(t_E) = E_n(t_E - \Delta t)$ -1 and $N_1(t_E) = 1$. Note that the logistic hatching rate is expressed as a function of the egg number $E_n$ rather than nymphs $N_1$. In fact, neglecting the Heaviside function, sinks and moulting, logistic hatching writes as $dN_1/dt = \varepsilon N_1(1 - N_1/K)$; because $E_{n0} = E_n(t = 0) = N_1(t) + E_n(t)$ must be satisfied at any time under these hypotheses, the substitution $N_1 = E_{n0} - E_n$ into the $dN_1/dt$ leads to $dN_1/dt = \varepsilon E_n(1 - E_n/K)$. Inclusion of the Heaviside function, predation on eggs and nymphys, and moulting results in Eq. (1) and Eq. (2) for logistic hatching.

Eq. (2) describes the rate of change of $N_1$ nymphs by logistic egg hatching, biotic and abiotic sinks, and moulting to $N_2$ nymphs, while Eq. (3) is the rate of change of $N_2$ nymphs by moulting from $N_1$ nymphs, biotic and abiotic sinks, and adult emergence[1]. A natural mortality rate of $N_1$ and $N_2$ individuals was not included explicitly as these were implicitly lumped into the rate $p$. Note that field sampling of $N_1$ and $N_2$ nymphs is normally performed by counting the number of individuals on 100 leaves (typically one leaf per plant). This non-destructive sampling does not require sinks in Eq. (2) and Eq. (3) but implies knowledge of the leaves distribution in order to allow model comparison with *in-situ* observations. In particular, the average number of nymphs sampled on 100 leaves can be written as

$$N_{1I}(t) = N_1(t)\frac{100}{f_{2y}n_{lP}P} ,$$
(6a)

$$N_{2I}(t) = N_2(t)\frac{100}{f_{2y}n_{lP}P} ,$$
(6b)

where $f_{2y}$ is the fraction of leaves near old wood (at least two years) of *V. vinifera* where *S. titanus* females lay eggs, $n_{lP}$ is the average number of leaves per plant, and $P$ is the number of plants in the vineyard. $N_{1I}$ and $N_{2I}$ are the actual quantities to be compared with experimental observations.

Eq. (4) is the adult rate of change due to emergence from $N_2$ nymphs, mortality, and biotic and abiotic sinks. Adult migration from and to the vineyard were not explicitly included. The last term on the right-hand side ($-y\rho_{YST}SI$) represents a sink due to population sampling. *S. titanus* adults are typically sampled with yellow sticky traps (YST) changed periodically (e.g., one or two weeks). YST sampling interferes with the population presence by removing free-living adults according to a characteristic function. This dynamical interference may only marginally affect the total *S. titanus* presence in the field but its sampling rate (*i.e.*, its kinetics) must be taken into account explicitly to allow for model comparison with *in-situ* YST measurements. To fulfil this requirement, the sampling rate $y$ per YST is made explicit in Eq. (4), with $\rho_{YST}$ the number of YST per unit surface area and $S$ the vineyard area. The average number of *S. titanus* adults $I_{YST}$ removed by one YST during its life time $\Delta t_{YST}$ is therefore

$$I_{YST}(t) = \int_{t-\Delta t_{YST}}^{t} yI(t^*)\,\mathrm{d}t^* ,$$
(6c)

and represents the actual quantity to be compared with experimental observations.

Finally, Eq. (5) describes the deposition rate of eggs in year $n$ that will overwinter to year $n + 1$, where $f_{MF}$ is the *S. titanus* adult sex ratio, and $t_L$ is the timing when adults become capable to lay eggs[2]. Although $f_{MF} = 0.5$ can be taken as a yearly average, $f_{MF}$ in *S. titanus* changes widely over the seasons (e.g., Bosio and Rossi 2001). Whereas logistic-like functions were used for $f_{MF}$ (e.g., Lessio et al. 2009), we define the time-dependent male-to-female sex ratio with the sigmoid function

$$f_{MF}(t) = 1 - \frac{1}{1+e^{-(t-a)/b}} = \frac{e^{-(t-a)/b}}{1+e^{-(t-a)/b}} ,$$
(7)

where $t$ is the time within year $n$, and $a$ and $b$ are empirical parameters with the dimension of time. Upon determination of $a$ and $b$, the number of adult males and females at any times of the year can be calculated as

$$I_M(t) = f_{MF}(t) \times I(t) ,$$
(8a)

$$I_F(t) = \left[1 - f_{MF}(t)\right] \times I(t) .$$
(8b)

---

[1] Note that the Heaviside function in Eq. (3) and Eq. (4) was not reported in the original presentation of the equations in Maggi et al., (2013).

[2] The delay in egg deposition $t_L$ was newly developed in this work and was not included in the original work in Maggi et al., (2013).

When passing from year $n$ to year $n+1$, the equations for eggs (current yearly number) prescribe $E_n(t + \Delta t) = E_{n+1}(t)$ and $E_{n+1}(t + \Delta t) = 0$, while the other stages are set to $N_1(t + \Delta t) = 0$, $N_2(t + \Delta t) = 0$, $I(t + \Delta t) = 0$.

Excluding migration from Eq. (4), the *S. titanus* population inventory within the vineyard of surface area $S$ at any times $t$ of the year must satisfy the population balance equation

$$E_{n0} = E_n(t) + N_1(t) + N_2(t) + I(t) + S\sum_m \rho_{YST} I_{YST}(t) +$$
$$+ \int_0^t p\left[E_n(t^*) + N_1(t^*) + N_2(t^*) + I(t^*)\right]dt^* + \int_0^t \delta I(t^*)dt^* \qquad (9)$$

where $m$ is the number of YST substitutions and $t^*$ is a dummy variable for time. Eq. (9) states that the initial number of eggs $E_{n0}$ in year $n$ must equal the total number of individuals present in the four stages at time $t$ plus the total number of individuals lost by $m$ YST sampling sequences, biotic and abiotic sinks, and natural mortality until time $t$.

The parameters $\varepsilon$, $\eta_1$, $\eta_2$, $\delta$, $t_E$, $t_{N1}$, $t_{N2}$, $t_L$, $\rho$, $p$, and $y$ in Eqs. (1)-(6), $f_{2y}$ and $n_{lp}$ in Eqs. (6), and $a$ and $b$ in Eq. (7) are physically measurable and are the basis to the mechanistic accounting of *S. titanus* populations proposed here.

## II.3 GOVERNING EQUATIONS OF FD ACQUISITION IN *S. TITANUS*

The state variables describing the presence of FD pathogen in *S. titanus* are the total number of adults $I$, the number of infected adults $I_i$ (including both latently infected and infective) and the number of infective adults $I_t$ (Figure 1). Accordingly, the number of healthy adults $I_h$ and latently-infected adults $I_L$ are secondary state variables that can be calculated as a function of $I$, $I_i$ and $I_t$. These must satisfy at any times the conservation equations

$$I_h(t) = I(t) - I_i(t), \qquad (10a)$$

$$I_L(t) = I_i(t) - I_t(t), \qquad (10b)$$

$$I(t) = I_h(t) + I_L(t) + I_t(t), \qquad (10c)$$

Analogously, *S. titanus* males and females belonging to healthy, infected and infective groups can be calculated using $f_{MF}(t)$ of Eq. (7).

The instantaneous rate of change in $I_i(t)$ and $I_t(t)$ at time $t$ can be written as

$$\frac{dI_i(t)}{dt} = \alpha D\frac{P_t(t)}{P(t)}I_h(t) - \left(\delta_i + p + y\rho_y S\right)I_L(t) - \left(\delta_t + p + y\rho_{YST}S\right)I_t(t), \qquad (11)$$

$$\frac{dI_t(t)}{dt} = H(t - L_I)\times \alpha D\frac{P_t(t-L_I)}{P(t-L_I)}I_h(t-L_I)e^{-(\delta_i + p + y\rho_{YST}S)L_I} - \left(\delta_t + p + y\rho_{YST}S\right)I_t(t), \qquad (12)$$

where $D$ is the insect mobility rate defining the number of plants visited per unit time, $P_t/P$ is the probability that a healthy *S. titanus* adult approaches an infected *V. vinifera* plant and $\alpha$ is the probability that a healthy insect becomes infected after feeding on that plant (e.g., Brauer et a., 2008; Sisterson, 2009). The second and third terms in Eq. (11) represent sinks as per Eq. (1) – (5) while the first term on the right-hand side of Eq. (12) represents the number of healthy insects that became infected at time ($t$-$L_i$) and have survived for a time equal to the latency $L_I$.

Bressan et al. (2005) showed that FD pathogen substantially affects *S. titanus* survival time. According to those observations, group-specific adult mortality rates were determined in Maggi et al., (2013) and were included in Eq. (11) and Eq. (12). The mortality rate $\delta$ in Eq. (4) can therefore be described as the weighted average

$$\delta = \frac{\delta_h I_h(t) + \delta_i I_L(t) + \delta_t I_t(t)}{I(t)}, \qquad (13a)$$

$$\delta_h = \delta_{hM} f_{MF} + \delta_{hF} \left( 1 - f_{MF} \right), \tag{13b}$$

$$\delta_i = \delta_{iM} f_{MF} + \delta_{iF} \left( 1 - f_{MF} \right), \tag{13c}$$

$$\delta_t = \delta_{tM} f_{MF} + \delta_{tF} \left( 1 - f_{MF} \right), \tag{13d}$$

where $\delta_h$, $\delta_i$ and $\delta_t$ are group-specific mortality rates.

Bressan et al., (2005) also reported about 50% decreased egg deposition in FD-affected *S. titanus* females. To include this effect explicitly in the governing equations of *S. titanus* life cycle under the effect of FD infection, the oviposition rate $\rho$ in Eq. (5) was written as

$$\rho = \frac{\rho_h I_h(t) + \rho_i I_L(t) + \rho_t I_t(t)}{I(t)}, \tag{14}$$

where $\rho_h$, $\rho_i$ and $\rho_t$ are group-specific oviposition rates[3]. No data currently exist on the mortality and oviposition rates of latently infected individuals, that is, whether all infected individuals have similar feedback on fitness regardless of symptoms; therefore, we assume $\delta_i = \delta_t$ and $\rho_i = \rho_t$ in Eqs. (13) and Eq. (14). Additionally, the oviposition rate $\delta_i = 0.5 \delta_h$ in Eqs. (13) was taken after experimental observation in Bressan et al., (2005).

## II.4 GOVERNING EQUATIONS OF FD EPIDEMICS IN *V. VINIFERA*

The state variables describing the presence of FD pathogen in *V. vinifera* are the total number of plants $P$, the number of infected plants $P_i$ (including both latently infected and infective) and the number of infective plants $P_t$ (Figure 1). Using these state variables, it is possible to calculate the secondary state variables

$$P_h(t) = P - P_{i,n}(t), \tag{15a}$$

$$P_L(t) = P_{i,n}(t) - P_t(t), \tag{15b}$$

$$P(t) = P_h(t) + P_L(t) + P_t(t), \tag{15c}$$

where $P_h$ and $P_L$ are the number of healthy and latently-infected (infected but not yet symptomatic) plants, respectively. When $P$ is constant as in typical perennial crops, the rate of change in number of FD-infected and infective plants at time $t$ of year $n$ can be written as

$$\frac{dP_{i,n}(t)}{dt} = \beta D \frac{P_h(t)}{P} I_t - r P_t(t), \tag{16}$$

$$\frac{dP_t(t)}{dt} = -r P_t(t), \tag{17}$$

where $D$ is the insect mobility rate, $P_h/P$ is the probability that an infected *S. titanus* adult moves to a healthy plant, $\beta$ is the probability that FD is transmitted from the infected insect to that plant, and $r$ is the plant recovery rate. Note that from year $n$ to year $n+1$ in Eq. (17), the number $P_t$ at the beginning of year $n+1$ equals $P_{i,n}$ at the end of year $n$ after recovery and implies that the latency in plants $L_P$ is at least equal to the overwinter time because symptoms of FD normally become visible the year after inoculation. While Eq. (16) does not require specific adjustments, the passage from year $n$ to year $n+1$ is numerically solved for Eq. (17) as

$$P_t(t + \Delta t) = P_{i,n}(t) - \Delta t \times r P_t(t). \tag{18}$$

---

[3] Note that this feedback of FD phytoplasma on *S. titanus* fitness was not included in the original formulation in Maggi et al., (2013)

FD-induced and natural plant mortality was not included here because removal of infected plants is a common practice in viticulture; specifically to this study, plant roguing was included as a practice to reduce FD infection sources as described in the following section.

## II.5 ACCOUNTING OF INSECTICIDE APPLICATIONS AND PLANT ROGUING

When insecticides are applied to decrease the presence of *S. titanus* vector, the governing equations must include suppression of a defined fraction of individuals at the time of insecticide application $t_{PST}$. The insecticide applications timing $t_{PST}$ may vary largely but in our applications this satisfies the relationship $t_{PST} > t_{N1}$ and is therefore assumed to suppress *S. titanus* individuals belonging to $N_1$, $N_2$ and $I$. The suppression efficiency $R_{PST}$ can vary largely depending on the active ingredient and application technique. For modelling purpose, insecticide applications were accounted for as an instantaneous suppression of individuals at time $t = t_{PST}$ and the fraction of surviving individuals was calculated as $(1-R_{PST})$. Using an explicit scheme, when $t = t_{PST}$, insecticide suppression was written as

$$N_1(t_{PST}) = (1 - R_{PST})N_1(t_{PST} - \Delta t) \tag{19a}$$

$$N_2(t_{PST}) = (1 - R_{PST})N_2(t_{PST} - \Delta t) \tag{19b}$$

$$I(t_{PST}) = (1 - R_{PST})I(t_{PST} - \Delta t) \tag{19c}$$

$$I_i(t_{PST}) = (1 - R_{PST})I_i(t_{PST} - \Delta t) \tag{19d}$$

$$I_t(t_{PST}) = (1 - R_{PST})I_t(t_{PST} - \Delta t) \tag{19e}$$

While applying this suppression to these state variables, the related secondary state variables defining gender are calculated accordingly using the $f_{MF}$ function of Eq. (7).

Plant roguing is applied to remove symptomatic plants, that is, those plans that can transmit the pathogen, and limit the exposure of healthy insect vectors to FD infection sources. If all infective plants are removed and replaced with healthy plants at a given time $t = t_{ROU}$ (normally during winter time), plant roguing can be accounted for as an instantaneous suppression of $P_t$ as

$$P_t(t_{rou}) = 0 \tag{20a}$$

$$P_{in}(t_{rou}) = P_{in}(t_{rou} - \Delta t) - P_t(t_{rou} - \Delta t) \tag{20b}$$

# III. MODEL TESTING PROCEDURE

## III.1. MODEL NUMERICAL SOLUTION

The governing equations for *S. titanus* dynamics, FD acquisition in *S. titanus*, and FD epidemics in *V. vinifera* are dynamically coupled and an analytical solution does not exist. A numerical solution was obtained *via* an implicit finite-difference integration technique. The model was implemented in the Matlab environment (Matlab 2011b). Numerical stability and mass (individual) conservation were verified in each simulation.

## III.2. PARAMETERS

The model presented here includes various parameters for *S. titanus* life cycle and FD epidemics in *V. vinifera*. Specifically, the parameters for *S. titanus* life cycle are the stage rates $\varepsilon$, $\eta_1$, $\eta_2$, $\delta_{hM}$, $\delta_{iM}$, $\delta_{hF}$, $\delta_{iF}$, $p$, $\rho_h$, $\rho_i$, YST sampling rate $y$, sex ratio parameters $a$, and $b$; and stage timing $t_E$, $t_{N1}$, $t_{N2}$ and $t_L$. Epidemiological parameters of two-way FD acquisition and transmission are the acquisition and transmission probabilities $\alpha$ and $\beta$, respectively, *S. titanus* vector mobility rate $D$ on *V. vinifera*, and *V. vinifera* recovery rate $r$.

## III.3. PARAMETER ESTIMATION OF *S. TITANUS* LIFE-CYCLE

Testing of the governing equations of *S. titanus* dynamics was carried out by parameter estimation and by parameter validation by statistical analysis.

Parameter estimation was performed by solving the inverse problem, i.e., using experimental values of the state variables to determine the parameters. This was carried out by minimizing the difference between observed and modeled data while modifying the parameters according to the Levenberg-Marquardt algorithm until residuals decreased below a tolerance (e.g., Levenberg 1944; Doerthy 2004). Thirteen independent data sets were used to the purpose of parameter estimation. These sets included data relative to egg hatching rate and adult longevity (Bressan et al. 2005), a comprehensive stage- and time-dependent table of yearly nymphs $N_1$ and $N_2$, and adult $I$ counting averaged over several Barbera vineyards in Piemonte (Bosco and Rossi 2001) and other vineyards in Piemonte and Friuli-Venezia Giulia regions (Bosco et al. 1997; Bianco et al. 2005), adult sex ratio (Bosco and Rossi 2001; Lessio et al. 2009), and a four-year *in-situ* nymph and adult counting used to estimate the egg deposition rate (Decante and van Helden 2006). In all parameter estimation sessions, the number of experimental points was always larger than the number of unknowns, thus preventing from mathematical indeterminacy.

Note that the calibration results presented in this report update the parameters presented in Maggi et al., (2013), and are the result of minor numerical improvements in the biological description of *S. titanus*, model solver and calibration procedure. Experimental data used for *S. titanus* life cycle parameter estimation are not tabulated but are represented along with modeling results in graphical form for ease of comparison.

## III.4. PARAMETER ESTIMATION OF FD EPIDEMICS IN *V. VINIFERA*

FD epidemics parameters could be estimated on various long-term survey data sets of FD presence in various vineyard sites carried out by the Regional Phytosanitary Service of Piemonte, Italy. Experimental observations are comprehensive from 1998 to 2006 and include yearly cumulative YST measurements of *S.*

*titanus* adult $I_y$; total, infected and latently infected plants $P$, $P_{i,n}$ and $P_L$; and number of replaced plants $P_{ms}$. Surveyed grapevines were Cortese (vineyard S2), Dolcetto (S6), Bonarda (S9) and Barbera (S13)[4] (see average values in Table 1). The monitored vineyard sections were about 500 m$^2$ and comprised about 200 plants. FD presence was detected in symptomatic plants by molecular diagnosis (Morone et al., 2007). *S. titanus* was sampled with 2 YSTs in each vineyard changed every 15 days from June 15 to October 10 every year. All vineyards received one insecticide treatment between May 30 and June 23, and a second treatment between July 3 and July 29.

These data were used to estimate the unknown parameters ($\alpha$, $\beta$, $D$, $r$, $p$) and initial conditions ($E_{n0}$, and $P_i = P_t$) using the observed yearly sequences of infected ($P_i$), latently-infected ($P_L$), infective plants ($P_t$), recovered plants ($P_r$), and sampled adult insects ($I_y$). The model was initialized to 10 years before the first year of experimental sampling; this spinup period was used to produce system dynamics over longer time scales and detect dynamics features (e.g., indefinitely exponential growth or decay, chaotic behaviours, numerical instabilities, etc.) that could not appear on the short term when initial conditions are to be chosen. Also in this case, parameter estimation was carried out by inverse problem solving. Testing the model on these data sets were particularly valuable because they included FD epidemics progression over nine years of field observations in various field sites.

**Table 1**. Experimental average count of infective plants ($P_t$), latently-infected plants ($P_L$), infected plants ($P_{in}$), substituted plants ($P_{ms}$), recovered plants ($P_r$), and yearly cumulative *S. titanus* adults ($I_y$) sampled with YSTs in various vineyards in Piemonte. Experimental data from 1998 to 2003 are from Morone et al., (2007), while data from 2004 to 2006 were provided by the Regional Phytosanitary Service of Piemonte, Italy. [a] calculated from tracking individual plants in two consecutive years.

|  |  | 1998 | 1999 | 2000 | 2001 | 2002 | 2003 | 2004 | 2005 | 2006 |
|---|---|---|---|---|---|---|---|---|---|---|
| Vineyards used for parameter estimation S2, S6, S9, S13 | $P_t$ | 33.0 | 97.0 | 86.8 | 88.3 | 33.0 | 39.3 | 47.8 | 39.8 | 28.8 |
| | [a] $P_L$ | 77.5 | 25.3 | 27.8 | 10.8 | 23.5 | 17.8 | 1.8 | 2.5 | - |
| | $P_{in}$ | 110.5 | 122.3 | 114.5 | 99.0 | 56.5 | 57.0 | 49.5 | 42.3 | - |
| | $P_{ms}$ | 0.0 | 18.0 | 13.0 | 14.5 | 8.3 | 0.0 | 1.3 | 2.3 | 2.8 |
| | $P_r$ | - | 13.5 | 34.5 | 25.5 | 34.3 | 16.8 | 9.3 | 8.8 | 4.5 |
| | $I_y$ | - | 173.0 | 40.3 | 8.0 | 6.5 | 1.8 | 1.3 | 2.8 | 3.8 |

### III.5. GOODNESS OF FIT

Goodness of fit relative to parameter estimation was measured with the correlation coefficient (R) and normalized root mean square error (NRMSE) defined as

$$\mathrm{R} = \mathrm{cov}(m, o) / \sigma_m \sigma_o \ ,$$

$$\mathrm{NRMSE} = \sqrt{\frac{1}{n} \sum_{i=1}^{n} \left( m_i - o_i \right)^2} \Big/ \left( \max\{o\} - \min\{o\} \right),$$

where *m* and *o* are the modeling and observation points.

---

[4] Here, we have used the label S followed by an identification number as from Morone et al., (2007).

## IV. EXPERIMENTS AND MODELLING OF *S. TITANUS* LIFE CYCLE

Selected results of a comprehensive model testing addressing parameter calibration of *S. titanus* dynamics are presented and discussed in this section.

### IV.1 *S. TITANUS* EGG HATCHING RATE

Bressan et al. (2005) measured the hatching rate $\varepsilon$ of eggs laid by 17 healthy and 22 FD-affected females after FD acquisition from broad bean and 43 days of incubation at 22-30 °C temperature. Their data showed that FD pathogen affected $\varepsilon$, the total number of eggs laid and their hatching timing. These observations were used to estimate $\varepsilon$ by solving Eq. (1) in both scenarios of eggs laid by healthy and FD-infected females. The value $p = 0$ d$^{-1}$ was used in Eq. (1) because the experiments were carried out in laboratory conditions in the absence of environmental sinks. Additionally, also the timing $t_E$ was determined from these experiments, but it could not be included in the analyses as this was not related to the yearly time frame of the univoltiine cycle of *S. titanus*.

The estimated values of $\varepsilon$ produced accurate results in both scenarios (Figure 2) and achieved R = 99.7% and NRMSE = 6.13% against experiments for eggs from healthy females, and R = 99.3% and NRMSE = 8.8% for eggs from FD-infected females, respectively (Table 2, column 1 and 2). We also noted that eggs laid by healthy females hatched slightly earlier than those laid by FD-affected females ($t_E$ = 18.8 d and 20.7 d, respectively, representing the time since beginning of observations), and that the average $\varepsilon$ was higher for eggs from healthy females. Note in Figure 2 that $t_E$ is the time when 1 egg hatches as describet in Section II.2. Overall, our modeling results complied with the conclusions put forth in Bressan et al. (2005). These values of $\varepsilon$ were further validated with independent data in the next sections.

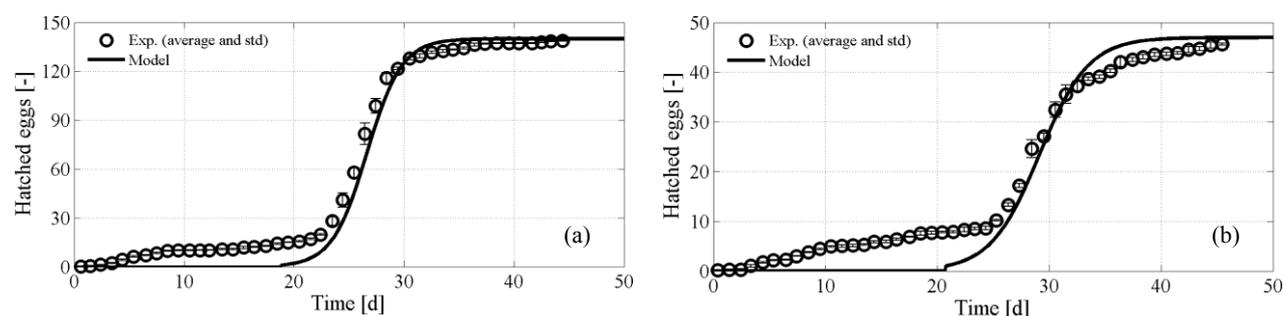

**Figure. 2.** Experimental and modeled rate of egg hatching by (a) healthy and (b) FD-affected *S. titanus* females. Experiments are from Bressan et al. (2005) and represent average and standard deviation of 3 replicates. Modeled data are from integration of Eq. (1) with the parameters summarized in Table 1, column 1 and 2.

**Table 2.** Summary of the parameters used to calculate the *Scaphoideus titanus* life cycle and population dynamics. Values in bold font are estimated from experiments. [a]Data from typical cultivar setting in Piemonte, Italy (Bosio and Rossi 2001). [b] Data from Bosco et al. (1997). [c] Data from Bianco et al. (2005). [d] Data from Decante and van Helden (2006).

| | | | Egg hatching | | Adult mortality | | | | Life cycle and population sampling | | | | | Sex ratio | Egg deposition | Statistics on *S. titanus* | | | | Epidemics |
|---|---|---|---|---|---|---|---|---|---|---|---|---|---|---|---|---|---|---|---|---|
| | | | Eq. (1) | | Eq. (4) | | | | Eq. (1) - Eq. (4), Eqs. (6) | | | | | Eq. (7) | Eq. (1)-Eq. (8) | | | | | Eq. (1)-Eq. (20) |
| | | | H | FD | H | | FD | | Regional | Site-specific | | | | | | | | | | |
| | | | | | M | F | M | F | Various sites | Brusnengo | Sostegno | Orcenico | Pasiano | | | Min. | Avg. | Max. | Std. | |
| | Column | | 1 | 2 | 3 | 4 | 5 | 6 | 7 | 8 | 9 | 10 | 11 | 12 | 13 | 14 | 15 | 16 | 17 | 18 |
| **Vine yard settings** | $S$ | $[\mathrm{m}^2]\times10^3$ | - | - | - | - | - | - | [a]10 | [b]3 | [b]2 | [c]5 | [c]5 | - | [d]14 | - | - | - | - | - |
| | $P$ | [-] | - | - | - | - | - | - | [a]3500 | [b]1050 | [b]700 | [c]1750 | [c]1750 | - | [d]6440 | - | - | - | - | - |
| **S. titanus parameters** | $\varepsilon$ | $[\mathrm{d}^{-1}]\times10^{-1}$ | **6.307** | **4.567** | - | - | - | - | **2.923** | **9.193** | **14.868** | **8.745** | **1.615** | - | **3.067** | 1.6 | 6.4 | 14.8 | 4.3 | 6.4 |
| | $\eta_1$ | $[\mathrm{d}^{-1}]\times10^{-2}$ | - | - | - | - | - | - | **9.248** | **4.506** | **8.054** | **14.21** | **14.56** | - | **8.290** | 4.5 | 9.8 | 14.5 | 3.9 | 9.8 |
| | $\eta_2$ | $[\mathrm{d}^{-1}]\times10^{-2}$ | - | - | 0 | 0 | 0 | 0 | **5.832** | **28.63** | **7.836** | **49.30** | **52.15** | - | **2.243** | 5.8 | 24.3 | 52.1 | 22.4 | 24.3 |
| | $\delta$ | $[\mathrm{d}^{-1}]\times10^{-2}$ | - | - | **4.02** | **1.60** | **23.34** | **8.87** | **3.138** | **10.19** | **16.40** | **3.810** | **2.771** | - | **6.297** | 1.6 | 8.0 | 23.3 | 6.9 | 8.0 |
| | $a$ | [d] | - | - | - | - | - | - | - | - | - | - | - | **200** | 200 | - | - | - | - | 200 |
| | $b$ | [d] | - | - | - | - | - | - | - | - | - | - | - | **12** | 12 | - | - | - | - | 12 |
| | $\rho$ | $[\mathrm{d}^{-1}]\times10^{-1}$ | - | - | 0 | 0 | 0 | 0 | - | - | - | - | - | - | **7.009** | - | 7.0 | - | - | 7.0 |
| | $p$ | $[\mathrm{d}^{-1}]\times10^{-3}$ | 0 | 0 | 0 | 0 | 0 | 0 | **21.52** | **6.023** | **5.545** | **9.986** | **24.56** | - | **5.999** | 5.5 | 11.3 | 24.5 | 7.4 | **7.352** |
| | $y$ | $[\mathrm{d}^{-1}]\times10^{-4}$ | - | - | - | - | - | - | **0.099** | **1.558** | **4.775** | **0.159** | **0.114** | - | **0.925** | 0.1 | 1.2 | 4.7 | 1.8 | 1.2 |
| | $\Delta t_{YST}$ | [d] | - | - | - | - | - | - | [a]15 | [b]7 | [b]7 | [c]10 | [c]10 | - | [d]7 | - | - | - | - | 7 |
| | $\rho_{YST}$ | $[\mathrm{m}^{-2}]\times10^{-3}$ | - | - | - | - | - | - | [a]5 | [b]6.7 | [b]5 | [c]5 | [c]5 | - | [d]4 | - | - | - | - | 4 |
| | $t_E$ | [d] | **18.8** | **20.7** | - | - | - | - | **134.5** | **116.7** | **105.5** | **146.3** | **115.7** | - | **103.6** | 105 | 120 | 146 | 16.7 | 120 |
| | $t_{N1}$ | [d] | - | - | - | - | - | - | **149.7** | **182.9** | **185.9** | **153.8** | **124.3** | - | **153.2** | 124 | 158 | 185 | 22.9 | 158 |
| | $t_{N2}$ | [d] | - | - | - | - | - | - | **175.1** | **191.1** | **188.6** | **187.5** | **180.5** | - | **194.1** | 175 | 186 | 191 | 7.0 | 186 |
| | $t_L$ | [d] | - | - | - | - | - | - | - | - | - | - | - | - | **204.1** | - | 204 | - | - | 204 |
| **V. Vinifera epidemics parameters** | $\alpha$ | $[-]\times10^{-2}$ | - | - | - | - | - | - | - | - | - | - | - | - | - | - | - | - | - | **0.659** |
| | $\beta$ | [-] | - | - | - | - | - | - | - | - | - | - | - | - | - | - | - | - | - | **0.144** |
| | $D$ | $[\mathrm{d}^{-1}]$ | - | - | - | - | - | - | - | - | - | - | - | - | - | - | - | - | - | **1.165** |
| | $r$ | $[\mathrm{d}^{-1}]\times10^{-4}$ | - | - | - | - | - | - | - | - | - | - | - | - | - | - | - | - | - | **8.574** |
| **Initial value** | $E_{n0}$ | [-] | **140** | **47** | - | - | - | - | 1000 | 3450 | 23306 | 38794 | 11766175 | - | **36893** | - | - | - | - | - |
| | N. exp. points | | 45 | 46 | 55 | 81 | 24 | 74 | 21 | 15 | 15 | 11 | 11 | 4 | 73 | | | | | 50 |
| | R% | | 99.71 | 99.35 | 97.5 | 86.04 | 95.77 | 99.11 | 94.69 | 94.16 | 97.86 | 93.13 | 97.99 | 97.69 | 97.02 | | | | | 93.50 |
| | NRMSE% | | 6.13 | 8.86 | 10.42 | 14.51 | 10.19 | 4.46 | 12.06 | 11.39 | 7.51 | 12.69 | 7.05 | 6.9 | 5.94 | | | | | 10.88 |

## IV.2 *S. TITANUS* ADULT LONGEVITY AND MORTALITY RATE

Bressan et al. (2005) recorded the longevity of healthy and FD-infected male and female adults emerged from 300 nymphs. Adults were caged on FD-infected broad bean for an acquisition period of 13 days, or on healthy broad-bean and were subsequently confined on young grapevine cuttings in plastic cages until death. FD presence was determined by PCR analyses (see details in Bressan et al. 2005). Equation (4) was used with these data to estimate the mortality rate $\delta$, with $\eta_2 = 0$ d$^{-1}$ (i.e., no moulting from $N_2$ nymphs), $p = 0$ d$^{-1}$ (i.e., no sinks), and $y = 0$ d$^{-1}$ (i.e., no YST sampling).

The numerical solution of Eq. (4) is represented in Figure 3 against experiments, while the estimated values of $\delta$ specific to each group are summarized in Table 2, column 3-6. Model fit was appreciable for FD-infected adults, but described healthy adult survival only approximately (Figure 3). Because the overall correlation coefficient between Eq. (4) and experiments ranged from 86.04% to 99.11%, with an average error in the range from 10.19% to 14.51%, first-order mortality was considered sufficiently accurate for the purpose of capturing large time scale dynamics of *S. titanus*, whereas different hypotheses of longevity models may be considered should on aim at the small time scale mathematical modeling of insect biological cycle.

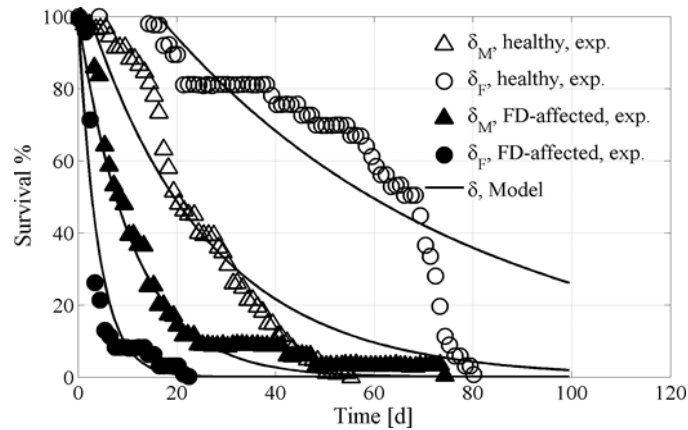

**Figure 3.** Experimental and modeled longevity of *S. titanus* males and female adults of healthy and FD-affected groups. Experiments are from Bressan et al. (2005). Modeled data are from integration of Eq. (4) with the parameters summarized in Table 2, column 3 to 6.

## IV.3 *S. TITANUS* AVERAGED POPULATION LIFE-CYCLE AND SAMPLING RATE

Estimation of *S. titanus* life-cycle and sampling parameters ($\varepsilon$, $\eta_1$, $\eta_2$, $\delta$, $t_E$, $t_{N1}$, $t_{N2}$, $p$, $y$) was carried out by solving Eq. (1)-Eq. (4) and Eq. (6) against data set earlier presented in Bosio and Rossi (2001). These data included a comprehensive stage-dependent table of nymphs and adult counting ($N_{1l}$, $N_{2l}$, and $I_{YST}$) from several Barbera vineyards in Piemonte in the years 1999 and 2000 during the months of June and July. Nymph sampling was conducted by counting the number of individuals on leaves near two-year old wood, where females lay eggs and 90% of immature are found. This sampling was carried out on 100 leaves per vineyard (1 leave per 100 plants or 2 leaves per 50 plants). Adult sampling was performed with YSTs changed every 14 days and vertically placed in the middle of the canopy at about 130-150 cm above the ground. In the same period, net sampling with 75 strokes per vineyard was carried out as a control. Insecticides were not used in the monitored vineyards (Bosio and Rossi 2001). Because experimental data reported the average relative presence, the empirical sampling fractions for the modeled nymph and adult presence were calculated as

$$N_{1l}\% = \frac{N_{1l}}{N_{1l} + N_{2l} + I_{YST}} 100$$

(10a)

$$N_{2l}\% = \frac{N_{2l}}{N_{1l} + N_{2l} + I_{YST}} 100$$

(10b)

$$I_y\% = \frac{I_y}{N_{1l} + N_{2l} + I_{YST}}100$$

(10c)

to allow for model-to-experiment comparison, with $N_{1l}$, $N_{2l}$, and $I_{YST}$ as in Eqs. (6). The number of leaves per plant $n_{lP}$ = 742 ± 63 was measured in a Barbera vineyard in Cocconato, Piemonte. From the same site inspections, the fraction of leaves per plant near two-year wood was estimated to be about 30%; hence, $f_{2y}$ = 0.3 $n_{lP}$ was used. Values $\rho_{YST}$ = 0.005 m$^{-2}$, $S$ = 10$^4$ m$^2$, and $P$ = 3500 were taken from cultivar setting. The initial number of eggs $E_{n0}$ = 1000 was fixed and was not relevant to the estimation process because data were expressed as relative percent values. The initial values $N_1(0) = N_2(0) = I(0) = 0$ were used for the remaining state variables.

The parameters corresponding to the best model-to-experiment fit are listed in Table 2, column 7, while Figure 4a shows a comparison between experimental and modeled $N_{1l}$, $N_{2l}$ and $I_{YST}$ (R = 94.69% and NRMSE = 12.06%). The calculated absolute presence of $E_n$, $N_1$, $N_2$ and $I$ complied with *S. titanus* populations typically recorded in Piemonte and elsewhere, which develop between spring and autumn months with no individuals surviving beyond October/November ($t > 300$ doy, Figure 4b).

Among the nine parameters used in Eq. (1)-Eq. (4) and Eq. (6), the times $t_E$, $t_{N1}$ and $t_{N2}$, predation $p$, and YST sampling $y$ strongly depended on the geographical location and various environmental factors, hence their significance was relative within the context of yearly-averaged *S. titanus* dynamics presented here. Rather, parameter estimation can be best appreciated in relation to $\varepsilon$, $\eta_1$, $\eta_2$, and $\delta$, as these parameters are intrinsic to *S. titanus*. Note that the egg hatching rate $\varepsilon$ about 50% lower than in laboratory experiments (Table 2, column 1 and 2) can be ascribed to the thermal variability in field conditions, while the mortality rate $\delta$ = 3.13 d$^{-1}$ complied with the values in Table 2, column 3-4 relative to healthy individuals.

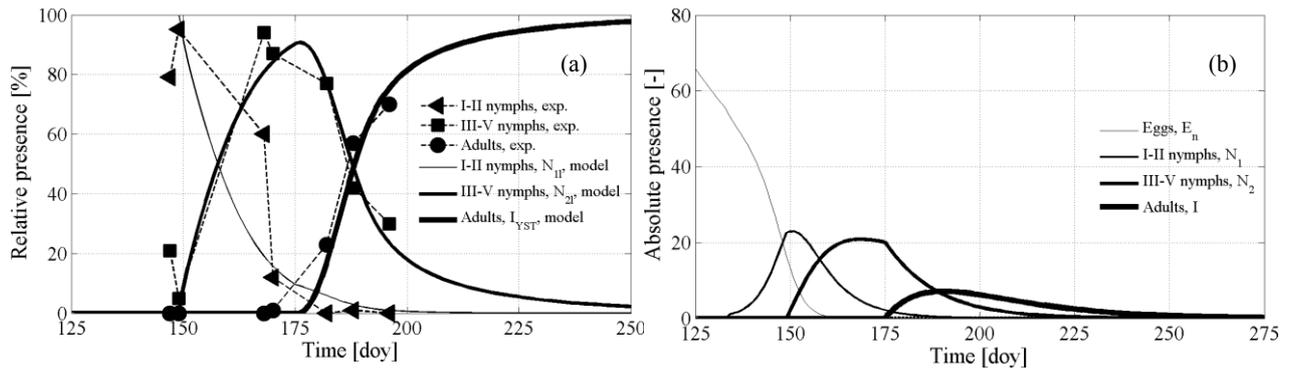

**Figure 4.** (a) Experimental and modeled relative presence of sampled I-II instar nymphs ($N_{1l}$), III-V instar nymphs($N_{2l}$), and adults ($I_y$) of *S. titanus* sampled in various vineyards in Piemonte, Italy. Experiments are redrawn from Bosio and Rossi (2001). (b) Absolute presence of eggs ($E_n$), I-II instar nymphs ($N_1$), III-V instar nymphs ($N_2$) and adults ($I$) corresponding to the sampled values of panel (a).

## IV.4 *S. TITANUS* SITE-SPECIFIC POPULATION LIFE-CYCLE AND SAMPLING

To highlight parameter sensitivity to site-specific conditions as a complement to average presence (previous section), life-cycle and sampling parameters $\varepsilon$, $\eta_1$, $\eta_2$, $\delta$, $t_E$, $t_{N1}$, $t_{N2}$, $p$ and $y$ where estimated against four independent YST measurements of adult presence collected in two Nebbiolo vineyards in Brusnengo and Sostegno, Piemonte, Italy (Bosco et al. 1997), and in two vineyards in Orcenico and Pasiano di Pordenone, Friuli-Venezia Giulia Region, Italy (Bianco et al. 2005). The surface $S$ of these vineyards were 3×10$^3$, 2×10$^3$, 5×10$^3$ and 5×10$^3$ m$^2$, respectively, with 1050, 700, 1750 and 1750 plants each, respectively, and with YST density $\rho_{YST}$ of 5×10$^{-2}$ to 6.7×10$^{-2}$ m$^{-2}$, respectively (see Table 2, column 8-11). The values $n_{lP}$ = 750 and $f_{2y}$ = 0.3 $n_{lP}$ were used as in previous section.

Comparison of observed and modeled adult sampling presence in Figure 5 shows that the governing equations could capture *S. titanus* dynamics in both qualitative and quantitative features. Over the four geographical scenarios, the model returned correlation coefficients R and error NRMSE in the range from 93.13% to 97.99% and from 7.05 to 12.69%, respectively (Table 2, column 8-11). We note here that all rates showed some variability as compared among the four sites and the average regional estimates. Similarly, the timing for initial hatching $t_E$, moulting $t_{N1}$ and emergence $t_{N2}$ had some variability.

These results suggest that geographical factors can affect specific parameters more or less substantially. Geographical factors may include temperature, solar radiation, air humidity, wind intensity, antagonist insects, etc. One way to account for parameter uncertainty linked to environmental factors is that of introducing a stochastic component in the equations presented here, aspect which is described in more detail in Maggi et al. (2013). Regardless of parameter uncertainty, the tests presented here and in the previous sections indicated that Eqs. (1)-(5) can be applied to describe *S. titanus* dynamics in independent vineyards with a relatively high accuracy.

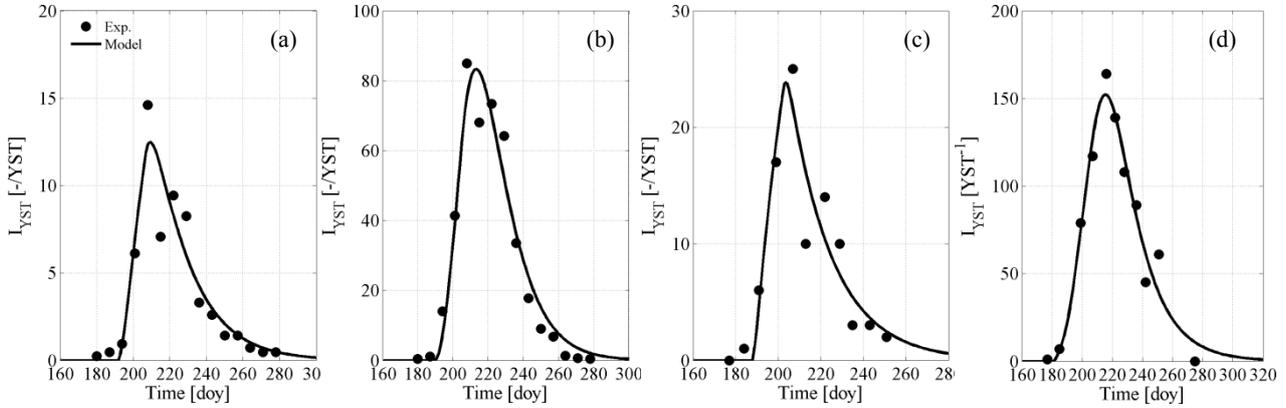

**Figure 5.** Experimental and modeled *S. titanus* adults sampled with YST in (a) Brusnengo, (b) Sostegno, (c) Orcenico and (d) Pasiano, Italy. Experiments in panels (a) and (b) are redrawn from Bosco et al., (1997), while experiments in panels (c) and (d) are redrawn from Bianco et al., (2005). Modeled data are from integration of Eq. (1) to Eq. (4), and Eq. (6) with the parameters summarized in Table 2, column 8 to 11.

## IV.5. *S. TITANUS* SEX RATIO

To account for gender diversity in the *S. titanus* governing equations, the adult sex ratio was measured between July and September 2000 in the Barbera vineyards described in Bosio and Rossi (2001).

Estimation of the parameters $a$ and $b$ of the male-to-female sex ratio $f_{MF}$ of Eq. (7) was performed by trial-end-error and resulted in $a = 200$ d and $b = 12$ d (Figure 6). With these parameters, Eq. (7) achieved a correlation coefficient R = 97.69% and error NRMSE = 6.9% against experiments (Table 2, column 12) and met the sex ratio from independent observations and modeling (e.g., Lessio et al. 2009).

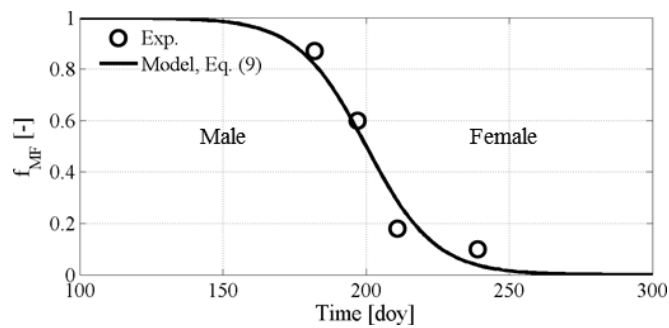

**Figure 6.** Experimental and modeled time-dependent sex ratio in *S. titanus*. Experiments are redrawn from Bosio and Rossi (2001) while modeled sex ratio is calculated from $f_{MF}$ in Eq. (9) with $a = 200$ d and $b = 12$ d.

## IV.6. *S. TITANUS* IN-SITU EGG DEPOSITION RATE AND MULTI-YEAR GENERATIONS

The insect sex ratio determined in Section IV.5, the egg deposition rate $\rho$ and timing $t_E$ (both unknown at this stage) are instrumental to determine the adult prolificacy and the population dynamics over multiple years. *S. titanus* nymph and adult counts from 1999 to 2002 in two vineyards near Bordeaux, France (Decante and van Helden 2006), were used to estimate $\rho$. These data were particularly valuable to understand how insect prolificacy affected population abundance in the following years in the absence of insecticide applications. For this test, Eq. (1)-(7) were solved with $S = 14 \times 10^3$ m$^2$, $P = 6440$, $\rho_{YST} = 4 \times 10^{-3}$ m$^{-2}$ and $\Delta t_{YST} = 7$ d as from *in-* *situ* conditions. The life-cycle and sampling parameters $\varepsilon$, $\eta_1$, $\eta_2$, $\delta$, $t_E$, $t_{N1}$, $t_{N2}$, $t_E$, $p$ and $y$ were estimated together with $\rho$ independently from previous estimations. Here, the values $N_1(0) = N_2(0) = I(0) = 0$ in year 1999 were used as initial conditions.

The modeling results showed some discrepancies in the timing when $I_{YST}$ increased (especially in year 2001 and 2002) and in the tail when $I_{YST}$ decreased (Figure 7). These discrepancies were probably caused by constant timing $t_E$, $t_{N1}$, $t_{N2}$ and $t_L$ in each year, whereas yearly variability related to environmental and climatic factors should be included. The corresponding absolute number of *S. titanus* individuals calculated in the same vineyards (Figure 8) show that the model captured relatively well the observed *S. titanus* populations and reached R = 97.02% and NRMSE = 5.94% (Table 2, column 13). The oviposition rate $\rho$ = 0.7 d$^{-1}$ (Table 2, column 13) appeared similar to laboratory observations $\rho \approx 1$ d$^{-1}$ (Bressan et al. 2005). In general, all other parameters were consistent with those estimated in the previous tests, whereas differences in sampling rate $y$ could be related to different traps used in Bosio and Rossi (2001) and in Decante and van Helden (2006), and also to other environmental factors.

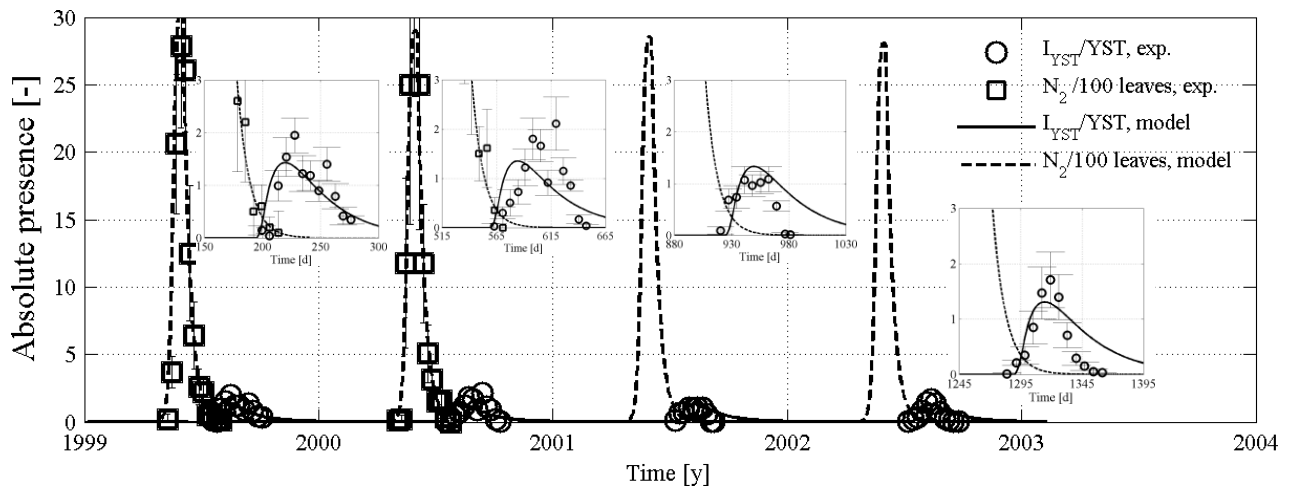

**Figure 7.** Experimental and modeled number of *S. titanus* nymphs and adults over 4 years measured in a vineyard near Bordeaux, France. (a)-(d) expanded view of sampled adult $I_y$ in every year. Experiments are redrawn from Decante and van Helden, (2006), while modeled data are from integration of Eq. (1)-(8) with the parameters in Table 2, column 13.

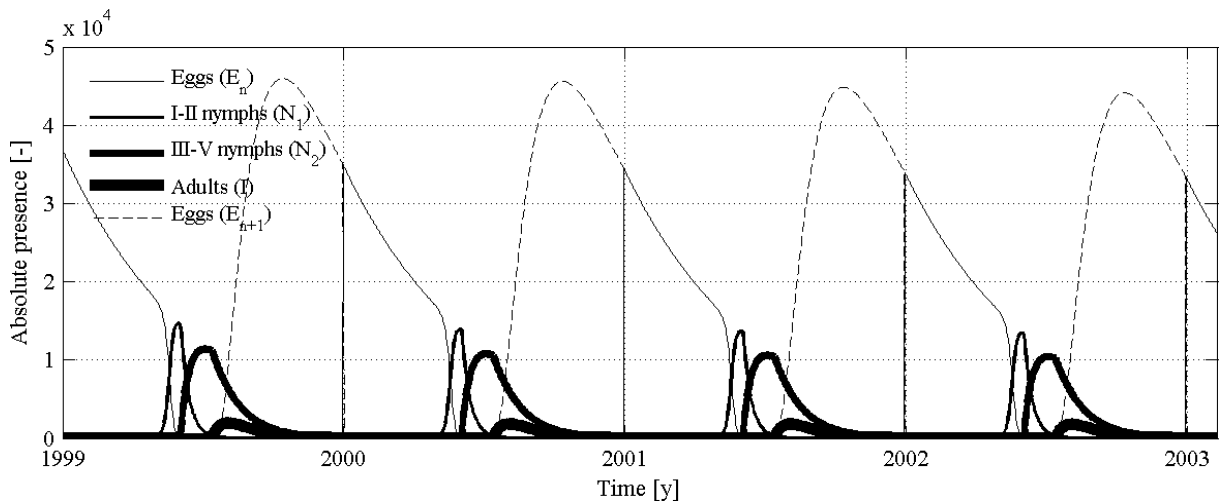

**Figure 8**. Modeled absolute presence of *S. titanus* individuals corresponding to the sampled nymphs and adults in Figure 7.

# V. EXPERIMENTS AND MODELLING OF IN-SITU FD EPIDEMICS IN *V. VINIFERA*

## V.1. ESTIMATION OF FD EPIDEMIOLOGICAL PARAMETERS

The governing equations describing *S. titanus* life cycle and dynamics, infection in *S. titanus*, and epidemics in *V. vinifera* were tested on data of FD epidemics recorded in sites S2, S6, S9 and S13 (see Table 1) to determine the insect mobility rate ($D$), acquisition and transmission probabilities ($\alpha$ and $\beta$), plant recovery rate ($r$), and site-specific sinks ($p$). The parameters for *S. titanus* life-cycle rates ($\varepsilon$, $\eta_1$, $\eta_2$, $\delta_{hM}$, $\delta_{lM}$, $\delta_{hF}$, $\delta_{lF}$, $\rho_h$, $\rho_l$), gender-specific parameters ($a$ and $b$), and timing ($t_E$, $t_{N1}$, $t_{N2}$, and $t_L$) were recalibrated in Section IV after Maggi et al., (2013) and are summarized in Table 2. Specifically, the average values of the above parameters were calculated from multiple estimation sessions and were used in this modeling exercise with no further adjustment (Table 2, column 18) with the exception of biotic and abiotic sinks rate $p$ as these are strongly correlated with in-situ conditions such as presence of pathogens, parasitoids and predators, and other environmental factors.

To test the governing equations in a coupled manner on the FD epidemics progression against field experiments we additionally constrained the model parameters when data were available. For example, insecticide applications were assumed to occur on day June 14 (163 doy) and July 18 (197 doy) as per recordings provided by the Regional Phytosanitary Service of Piemonte, Italy. Site observations of *S. titanus* presence showed that the population decreased by about 64.7% each year relative to the previous year. Assuming that observed decrease was exclusively caused by insecticide treatments, we inferred that each application had a suppression efficiency of about 40.6%. However, considering also abiotic and biotic sink $p$, a lower suppression efficiency is likely; on the basis of this consideration, $R_{PST} = 0.30$ was used for each insecticide application as per Eqs. (19).

The latency $L_I = 30$ d in *S. titanus* was used after our laboratory observations and other observations (e.g., Caudwell et al., 1970; Bressan et al., 2006), while for *V. vinifera*, $L_P$ was adjusted to result in symptoms appearing the year after as in Eq. (18). Infective plants removed because of death or heavy symptoms ($P_{ms}$, Table 1), were replaced with healthy plants and were accounted for by decreasing $P_t$ and $P_{i,n}$ by $P_{ms}$ at the end of each year as proposed in Eqs. (20). Finally, values $N_1(0) = N_2(0) = I(0) = 0$ for *S. titanus* were used as initial conditions in year 1996, while the initial egg presence $E_{n0}(0)$ was estimated as any other unknown parameter.

The parameters estimated in this test are summarized and highlighted in bolt font in Table 1, column 18. The FD acquisition probability $\alpha = 0.0065$ and transmission probability $\beta = 0.144$ suggest that FD transmission to plants was more efficient than acquisition from plant. Because more inoculum can be delivered with insect saliva than can be acquired from phloem sap, estimation of $\alpha$ and $\beta$ values was deemed sound. These results also suggest that a given amount of inoculum has greater impact in infecting plants than in infecting feeding insects. This idea, proposed by Purcell (1982), seems reasonable if we think that phytoplasmas are injected directly into the sieve tubes with vector saliva, while phytoplasmas acquired by feeding insects have to overcome anatomical barriers in order to reach the salivary glands and be transmitted.

The estimated insect mobility rate $D = 1.165$ plants visited each day appeared to be reasonable within this context and matched values of $D$ ranging between 1.16 and 1.4 plants per day obtained by the authors on another deltocephalinae leafhopper, *M. quadripunctulatus*, in controlled experiments inside climatic chambers (data not published). Note, however, that field observations of how frequently *S. titanus* moves from one plant to another do not currently exist for comparison.

A plant recovery rate $r = 8.54 \times 10^{-4}$ d$^{-1}$ suggested that the average duration of FD disease in one infected plant lasted for approximately 3 to 4 years. This estimate takes into account possible FD re-inoculation during the period when plant is infected, and suggests that recovery may be faster in the absence of new inoculations. In fact, preliminary (unpublished) studies on infected Barbera grapes in vineyards treated with insecticides since several years and with very low vector populations have shown that recovery may occur in period of time comparable to this estimate. Overall, however, specific biological (e.g., grape variety) and environmental factors (e.g., climatic conditions, water and nutrient availability, biotic and abiotic stresses) possibly contributing to plant recovery, are still unknown and therefore, final and general conclusions cannot be drawn on this aspect.

The modeling results in Figure 9 represent daily (gray) and end-of-year (black) values of the state variables. It is interesting to note that the modeled FD epidemics in plants (Figure 2a, b, c, d and e) captured the experimental FD progression curve both in qualitative terms (e.g., the overall trends and the curve position within the experimental standard deviation) and quantitative terms (R = 93.5% and NRMSE = 10.88%, see Table 2, column 18). The absolute number of sampled adults was monotonically decreasing over time and complied with the decreasing population observed by sampling (Figure 2e). The matching remained relatively accurate over 9 years of experimental observations and suggested the model be reliable to perform predictions over the long term. We note that this exercise of comparison between epidemiological modeling and recordings is not common practice, but we recommend that this should be attempted to provide evidence of model reliability and validity in all cases where data are available.

An analysis of *S. titanus* population showed a year-after-year increase in all *S. titanus* stages before 1998 and a subsequent decrease as a result of insecticide applications after 1998 (Figure 10a), which reduced the number of $N_1$, $N_2$ and $I$. Of the total number of *S. titanus* individuals, an increasing number underwent FD infection in the first years of the FD outbreak, while insecticide application resulted in a decrease of infected and infective *S. titanus* after 1999 (see $I_i$ and $I_t$ in Figure 10b). Surprisingly, only a very small number of *S. titanus* adults was infective and caused the epidemics in plants. These results are interesting in that the FD outbreak in *V. vinifera* reached a peak between 1998 and 2000, when also the number of mobile FD infective *S. titanus* adult showed the highest presence (Figure 10b).

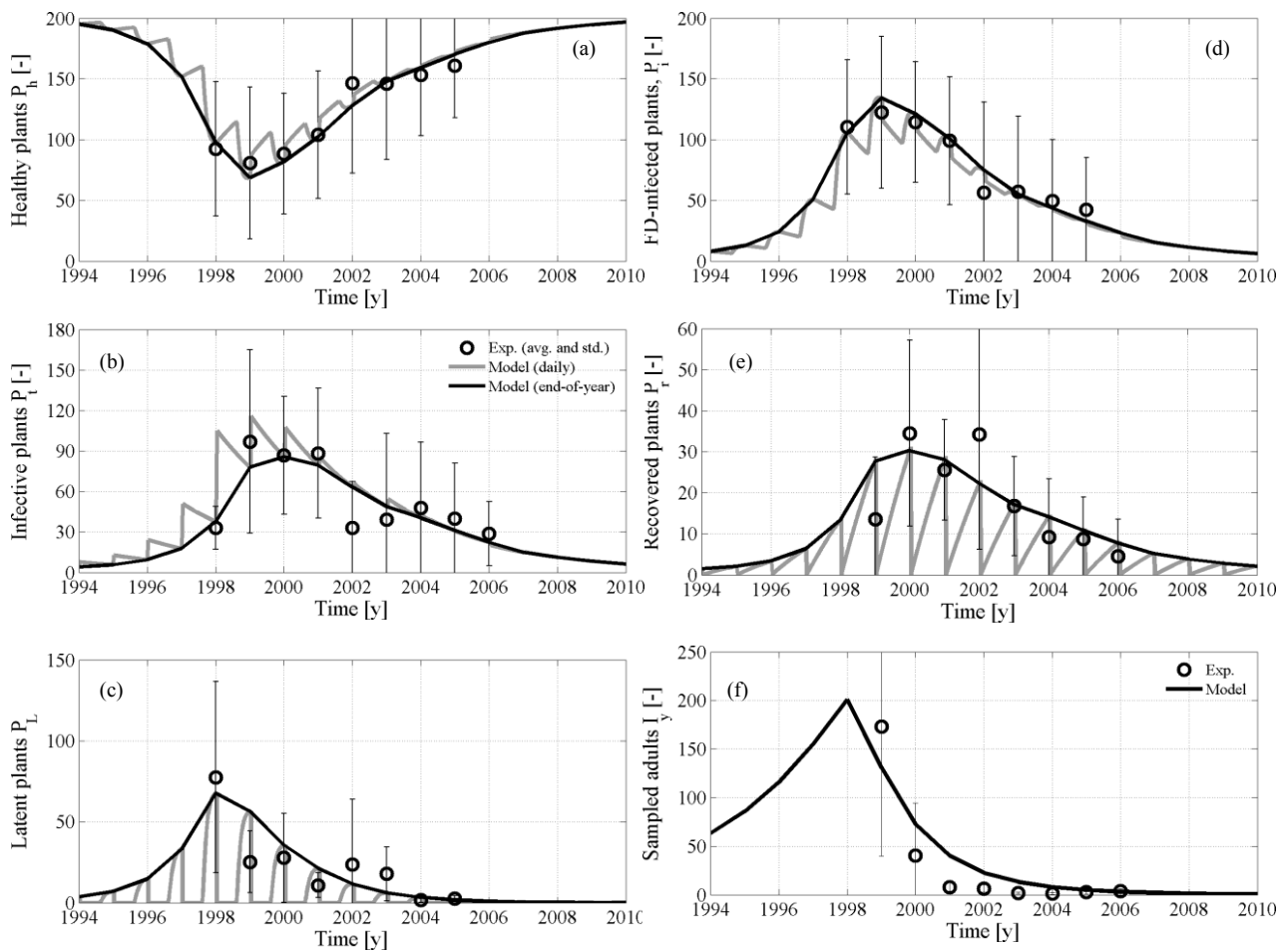

**Figure 9**. Experimental (circles) and modelled (lines) healthy plants ($P_h = P - P_i$) infective plants ($P_t$), latently-infected plants ($P_L$), infected plants ($P_{in} = P_t + P_L$), recovered plants ($P_r$) and *S. titanus* sampled adults ($I_y$). Experiments were averages over vineyards S2, S6, S9, S13 in Table 1 from Morone et al., (2007). Modelled values were calculated using parameters in Table 2, column 18. Gray and black curves represent daily and end-of-year values, respectively.

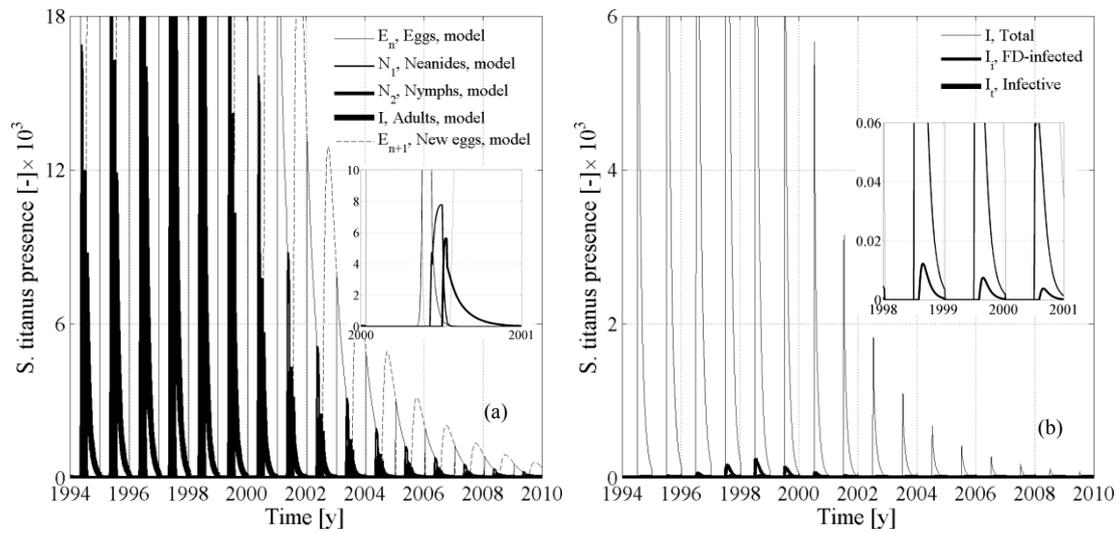

Figure 10. (a) modelled daily presence of *S. titanus* stages during the FD epidemics of Figure 9. (b) modelled daily total presence of *S. tianus* adults ($I$), infected adults ($I_i$) and infective adults ($I_t$) during the FD epidemics of Figure 9.

# VI. SUMMARY AND BRIEF DISCUSSION

## VI.1. ON THE BIOLOGICAL CYCLE OF *S. TITANUS*

We have presented in Section II.2 a continuous-in-time five-stage and size-dependent model of *S. titaus* life cycle. The model allowed to determine values of rate and timing parameters of the biological cycle of *S. titanus* using independent data sets. Comparison of parameters suggested these to be reliable in the ranges of standard deviation reported in the statistics of Table 2, while analysis of modeling results indicated that the governing equations for *S. titanus* were robust and could be applied in a variety of environmental settings. The model can therefore be useful in analyzing *S. titanus* population dynamics in that concerns with salient life stages occurring during the yearly time scale and beyond. In additions, the set of parameters summarized in Table 2 represents a unique collection of parameter values that can be used when biological data relative to *S. titanus* are not directly available and allow for further parameter and model verification. Minor changes were implemented in this model as compared to the original one presented in Maggi et al. (2013): this included the accounting of the timing $t_E$, which represents the time *S. titanus* adults require to be able to lay eggs that will overwinter to hatch the year after, and the effect of FD pathogen on group-specific egg deposition rate $\delta$. These new implementations required recalibration of the model, which led to changes in the estimated parameters. The newly estimated parameters are reported in Table 2 and show that only minor variations were required for the model to retain a high accuracy in replicating experimental data.

Overall, the model was developed on assumptions, which are discussed as follows. The model used for adult mortality was based on first-order kinetics. Although this is a typical approach in modeling biological species (e.g., Allen 2003; Sisterson 2009), the survival in healthy *S. titanus* adults recorded in Bressan et al. (2005) rather suggested that logistic mortality could be used. We note however that *S. titanus* populations found in vineyards affected by FD may have mixed healthy and FD-infected individuals. In these circumstances, a mortality rate averaged by gender and infected individuals can reasonably be considered to range within $1 \times 10^{-2}$ d$^{-1}$ and $20 \times 10^{-2}$ d$^{-1}$ depending on the density of specific groups and environmental factors. The values of $\delta$ estimated in this work described therefore those of in-situ populations within a tolerance. This is the case of mortality rates in the range from $2.9 \times 10^{-2}$ d$^{-1}$ to $16.4 \times 10^{-2}$ d$^{-1}$ determined from data in Figure 5 (Table 2, column 7-11), which averaged gender-specific, and FD-affected and healthy groups. Although a high variability, the model was able to capture with relatively high accuracy the observed adult dynamics over time.

As already noted in Bosio and Rossi (2001) and in Maggi et al. (2013), adults normally begin laying eggs only after a time lag from emergence and, presumably, the oviposition rate $\delta$ may change over time. In the updated model presented here, we have explicitly included a time lag by means of the parameter $t_E$ mentioned above. We have however maintained the first-order oviposition kinetics; while this may not fully describe the actual oviposition rate in field conditions, it was shown to be reliable within the model structure presented here as the population dynamics estimated over multiple years in Figure 7 showed the model to be capable of capturing the *S. titanus* dynamics well.

Thirteen independent parametric estimations presented here allowed us to analyze parameter variability. On the basis of these results, a Monte Carlo stochastic sensitivity analysis proposed in Maggi et al. (2013) showed that *S. titanus* life-cycle was more susceptible to timing parameters than rate parameters. Parameter validation was performed by a statistical analysis of the values obtained by parameter estimation over each individual data sets. A parameter which showed low standard deviation as compared to its range of variability (minimum and maximum values) was considered reliable for modeling purpose within a level of tolerance. In the statistics of Table 2, column 14-17, it appeared that substantially all parameters could be deemed reliable. Parameter variability within a reliable range was however associated to a level of parameter uncertainty linked to intrinsic stochasticity of environmental and biological factors as well as to model uncertainty such as biological processes not explicitly represented in the proposed equations.

Understanding model sensitivity to various parametric groups may have important implications toward *S. titanus* presence and FD disease epidemics; these analyses suggested that any changes in the actual value of these parameters may represent a means to reduce *S. titanus* presence and reduce potential risks of FD outbreaks. Hypotheses along this line were put forth in Rigamonti et al. (2011) on the basis of mechanistic modeling of *S. titanus* timing, yet to date no existing mathematical model of *S. titanus* has been developed at the level of the one presented here, neither a case study of pesticide control validated by field data is currently available in the scientific literature. Insecticide applications was included in the model proposed here by means of an instantaneous suppression of a fraction of nymphs and adults, and showed to be particularly effective in reducing the number of eggs laid in each year, hence the number of adults in the next years (excluding migration). Model interrogation and simulation results in Maggi et al. (2013) can aid in designing and improving the effectiveness of these operations by allowing us to select their optimal timing, frequency and lapse.

In that concern with the biological life-cycle of *S. titanus*, various aspects can be introduced to improve the model descriptive capability. The most important one is the coupled climatic and meteorological effects on *S. titanus* dynamics; for example, temperature has a major impact on insect life cycle and mobility (e.g., Honék 1996; Bressan et al., 2006; Chuche and Thiéry 2009), although solar radiation, wind, and precipitations may also play a role (Rigamonti et al. 2011). These factors should explicitly be accounted for in a degree-day based model of *S. titanus* development as well as in empirical expressions of the rate constants $\varepsilon$, $\eta_1$, $\eta_2$, $\delta$, and $\rho$. Factors affecting *S. titanus* dynamics such as temperature, precipitation, solar radiation, wind, antagonist insects, etc., are not explicitly included in the present work but estimation of the above parameters takes them into account in an implicit manner. In fact, part of the parametric variability highlighted by the standard deviation in Table 2, column 17, can be linked to the effect of factors not explicitly included in our mathematical model such as temperature.

## VI.2. ON THE COUPLED *S. TITANUS* - *V. VINIFERA* DYNAMICS RELATIVE TO FD EPIDEMICS

The governing equation of FD epidemics in Section II.2 to Section II.4 are, to the best of the authors' knowledge, the first dynamically coupled model of the biological life cycle of FD vector *S. titanus*, host plant *V. vinifera*, and FD pathogen development in both vector and plant. The model was tested on FD epidemics observed in vineyards in Piemonte, Italy, where a large number of control variables were recorded over several years. This is the first attempt that an insect-borne plant disease epidemiological model is tested on real data. We believe that this capability is uniquely made possible by the fact that a fully mechanistic life-cycle description of the insect vector *S. titanus* is used. Earlier models of insect-borne plant disease (Jeger et al., 2004; Madden et al., 2000; Madden and van den Bosch 2002; Nakazawa et al., 2012) have been proposed in the scientific literature but never have these been tested or validated on real epidemics because they miss description of the vector dynamics since they only include the adult stage of the vector insect. An exception to this gap is the work presented in Ferriss and Berger (1993), who used the number of infected plants over time as a metric to estimate and validate parameters in a stochastic framework.

The model presented in this work and the results demonstrated that the coupled pathogen-vector-host dynamics could describe and predict the epidemics progression curve with high accuracy within the observed standard deviation, achieving correlations with experiments R = 93.5% and residuals NRMSE = 10.88%. The matching between experiments and model simulations is encouraging and although epidemics parameters were not validated on independent tests in this instance, the results are promising and let us believe that additional data of the same density and quality may be used to further test the robustness of the coupled model.

As already discussed for the biological life cycle of *S. titanus*, and because *S. titanus* dynamics is clearly strongly correlated to the FD epidemics spreading, climatic and meteorological effects on FD epidemics could be included to improve model robustness. These factors should explicitly be accounted for in empirical expressions of the rate constants $\varepsilon$, $\eta_1$, $\eta_2$, $\delta$, $\rho$, and $D$, whereas it is likely that the epidemiological parameters $\alpha$ and $\beta$ are not affected.

With the proper numerical tools, the model presented and tested here for FD epidemic can be used to predict the effectiveness of mitigation strategies against FD epidemics spreading. For example, scenarios of insecticide applications can be designed within this model and optimized as a function of the number and time of applications either repeating the same practice over multiple years or adaptively changing application number and timing according to real-time reading of FD in-situ presence of *S. titanus* adults and *V. vinifera* plants. Analogously, scenarios of plant roguing can be designed and optimized using the model presented here as a function of the fraction of plants substituted and the economical implication in the reduced production and field management operations. Additional scenarios can be included where optimization is performed on combinations of insecticide applications numbers and timing, and roguing fraction. Note that roguing is normally applied to all symptomatic plants, while it is possible to consider that only a fraction of them may be substituted with healthy plants after economic considerations. These types of analyses can be extended to more scenarios and various agronomic situations, where the model can be used to inquire possible outcomes and help decision makers in planning short- and long-term strategies to control plant diseases in agriculture.

## VI.3. A PATHWAY FOR FUTURE RESEARCH

Among the modeling improvements discussed in Section VI.1 and VI.2, an aspect that deserves particular attention is the spatial component of *S. titanus* presence and FD-affected plants within a field, and the resulting time-space dependent epidemics progression. As for the vast majority of insect models, the model presented here is zero-dimensional, that is, it does not incorporate any spatial dependency of the system state variables, whereas spatial processes such as conditions at the boundaries as well as space-dependent insect mobility and migration may play a role in FD epidemics (Decante and van Helden 2006; Lessio et al. 2004). For example, stochastic random-walk processes are known to possibly describe biological processes of movement of individuals within a population. These are characterized by a travel distance defined by a probability density function (normally a negative exponential or a Gaussian function) and by a uniform distribution in the travel direction. Random walk processes may be potentially used to implement space-time stochastic descriptions of epidemics progressions and support theoretical epidemiology analyses, comparison with experimental data, and decision making. To date, spatial description of in-situ epidemics is very limited and could be a research line of potentially high impact in epidemiology.

## VII. CONCLUSIONS

This report summarises a comprehensive work aimed at developing and testing a mechanistic model of Flavescence Doree (FD) phytoplasma epidemcis in *V. vinfera* transmitted by the leafhopper *S. titanus*. FD epidemics were described by coupling *S. titanus* population dynamics and FD pathogen cycle in both *S. titanus* and in *V. vinifera*. *S. titanus* was described using a five-stage size-dependent model based on an earlier approach presented in Maggi et al. (2013), here improved in the description of adult oviposition kinetics. *S. titanus* rate and timing parameters were therefore re-assessed against a wide range of experimental data already used in the earlier parameter estimation exercise in Maggi et al., (12013) and updated values were presented in this report. Insects and plants were divided into subgroups describing healthy, infected, and infective individuals as primary system variables, while secondary variable describing latently-infected and healthy individuals were defined by linear dependence on the primary system variables.

The epidemics spreading in insects and plants were described by ordinary differential equations that scaled the rate of occurrence of new infections with a two-way probability of FD acquisition and transmission between insects and plants. The parameters used in the epidemiological module of the model were calibrated against various data of FD epidemics acquired in the last decade in Piemonte, Italy.

Tests on experimental data of the biological cycle of *S. titanus* and of FD in-situ epidemics showed that the model was overall highly accurate in predicting both short- and long-term dynamical features, and was robust in that it achieved high accuracy in independent data sets. The mathematical framework presented here let us therefore believe to be a beneficial tool when interrogated to provide scenarios of possible evolution and control strategies of insect-borne plant diseases in agriculture. This also sets the platform for further research initiatives, possible tackling an approach to describe the space-time epidemics progression.

## VIII. ACKNOWLEDGEMENTS

The authors are indebted with Dr. Chiara Morone and Dr. Paola Gotta, Servizio Fitosanitario, Regione Piemonte, Torino, Italy, for providing the authors with experimental data, technical support, and expertise. This work was funded by the Regione Piemonte, Italy, under the grant CIPE "Adoption of a multidisciplinary approach to study the grapevine agroecosystem: analysis of biotic and abiotic factors able to influence yield and quality". F. Maggi was partly supported by the 2010 Lagrange Fellowship, Fondazione CRT, Torino, Italy.

## IX. LIST OF SYMBOLS

| | | |
|---|---|---|
| $\alpha$ | [-] | probability of plant-to-insect FD transmission |
| $\beta$ | [-] | probability of insect-to-plant FD transmission |
| $\delta$ | $[T^{-1}]$ | insect mortality rate |
| $\delta$ | $[T^{-1}]$ | egg hatching rate |
| $\eta_1$ | $[T^{-1}]$ | moulting rate from I-II instar to III-V instar |
| $\eta_2$ | $[T^{-1}]$ | adult emergence rate from III-V instar |
| $\rho$ | $[T^{-1}]$ | egg deposition rate per female per unit time |
| $\rho_{YST}$ | $[m^{-2}]$ | traps per square metre |
| $D$ | $[T^{-1}]$ | insect mobility rate |
| $\Delta t_{YST}$ | $[T]$ | YST turnover time |
| *a, b* | [ d ] | sex-ratio parameters |
| $f_{FM}$ | [-] | male-to-female sex ratio |
| $I$ | [-] | total number of insects |
| $L_I$ | $[T]$ | latent period in insects |
| $L_P$ | $[T]$ | latent period in plants |
| $P$ | [-] | total number of plants |
| $p$ | $[T^{-1}]$ | predation rate |
| $r$ | $[T^{-1}]$ | plant recovery rate |
| $t_E$ | $[T]$ | time of egg hatching |
| $t_{N1}$ | $[T]$ | time of moulting |
| $t_{N2}$ | $[T]$ | time of emergence |
| $t_L$ | $[T]$ | time of oviposition |
| $t_{PSD}$ | $[T]$ | time of pesticide application |
| $t_{ROU}$ | $[T]$ | time of plant roguing |
| $y$ | $[T^{-1}]$ | YST sampling rate |

| | |
|---|---|
| NRMSE | Normalized Root Mean Square Error |
| R | Correlation coefficient |
| $h$ | subscript for healthy individuals |
| $i$ | subscript for infected individuals |
| $L$ | subscript for latently-infected individuals |
| $ms$ | subscript for substituted plants |
| $n$ | subscript for year |
| $t$ | subscript for infective individuals |

$y$     subscript for sampled insect
$M$     subscript for male individuals
$F$     subscript for female individual

# X. REFERENCES

Allen, L.J.S,. (2003), An introduction to stochastic processes with applications to biology, Pearson/Prentice Hall - Mathematics, pp. 385.

Anderson R.M., May R.M., (1979), Population biology of infectious diseases: Part I. Nature 280, 361-367.

Arnaud, G., Malembic-Maher A., Salar P., Bonnet P., Maixner M., Marcone C., Boudon-Padieu E., and Foissac X., (2007). Multilocus Sequence Typing confirms the close genetic interrelatedness of three distinct Flavescence Dorée phytoplasma strains clusters and group 16SrV phytoplasmas infecting grapevine and alder in Europe. Appl.  Environ. Microbiol. 73(12):4001-4010.

Bailey N.J.T., (1975), The Mathematical Theory of Infectious Diseases and its Applications, London, UK: Griffin.

Belli G., Fortusini A., Osler R., Amici A., (1973), Presenza di una malattia del tipo "Flavescence dorée" in vigneti dell'Oltrepò pavese. Riv. Patol. Veg. 9 S. IV (suppl):51-56.

Bellomo C., Carraro L., Ermacora P., Pavan F., Osler R., Frausin C., Governatori G., (2007), Recovery phenomena in grapevines affected by grapevine yellows in Friuli Venezia Giulia. First international phytoplasmologist working group meeting, Bologna, Italy, 12-15 November 2007. Bulletin of Insectology 60(2), 235-236.

Bertin S., Guglielmino C.R., Karam N., Gomulski L.M., Malacrida  A.R., Gasperi G., (2007), Diffusion of the Nearctic leafhopper Scaphoideus titanus Ball in Europe: a consequence of human trading activity. Genetica 131(3): 275-285.

Bianco, P.A., Loi N., Martini M., and Casati P., (2005), Flavescenza dorata, Quaderno Arsia 3:75-109. (in Italian).

Biere A., Bennett A.E., (2013), Three-way interactions between plants, microbes and insects. Functional Ecology 27, 567–573 doi: 10.1111/1365-2435.12100

Bosco D., Alma A., and Arzone A., (1997), Studies on population dynamics and spatial distribution of leafhoppers in vineyards (Homoptera: Cicadellidae). Ann. appl. Biol. 130:l-11.

Bosco D., Marzachi' C., (2011), Flavescenza Dorata in Barbera e Nebbiolo: incidenza, risanamento e suscettibilita' al patogeno. Protezione delle Colture, 21-23.

Bosco D., D'Amelio R., (2010), Transmission specificity and competition of multiple phytoplasmas in the insect vector. In "Phytoplasmas" (P.G. Weintraub and P. Jones Eds.), CAB International, Wallingford, UK, pp. 293-308.

Bosio G., Rossi A., (2001), Ciclo biologico in Piemonte di Scaphoidues titanus, L'Informatore Agrario, 21/2001, 75-78.

Brauer F., Van den Driessche P., Wu J., (2008), Mathematical epidemiology, Springer-Verlag Berlin Heidelberg.

Bressan A., Girolami V., Boudon-Padieu E., (2005), Reduced fitness of the leafhopper vector Scaphoideus titanus exposed to Flavescence dorée phytoplasma. Entomologia Experimentalis et Applicata 115, 283 –290.

Bressan A., Larrue J., Boudon-Padieu E., (2006), Patterns of phytoplasma-infected and infective Scaphoideus titanus leafhoppers in vineyards with high incidence of Flavescence dorée. Entomologia Experimentalis et Applicata 119, 61–69.

Carle P., and Moutous G., (1965), Observation sur le mode de nutrition sur vigne de quatre espèces de cicadelles. Annls. Epiphyt., 16, 333-354 (in French).

Caudwell A., (1957), Deux années d'études sur la Flaescence dorée, nouvelle maladie grave de la vigne. Annales de l'Amelioration des Plantes 4, 359-363.

Caudwell A., Kuszala C., Bachelier J.C., Larrue J., (1970), Transmission de la Flavescence dorée de la vigne aux plantes herbacées par l'allongement du temps d'utilisation de la cicadelle Scaphoideus littoralis BALL et l'étude de sa survie sur un grand nombre d'espèces végétales. Ann. Phytopathol. 2(2), 415-428.

Caudwell A., and Larrue J., (1986), La Flavescence dorée dans le Midi de la France et dans le Bas-Rhône. Progrès Agricole et Viticole 103(22): 517-523.

Caudwell A., Boudon-Padieu E., Kuszala C., Larrue J., (1987), Biologie et étiologie de la Flavescence dorée. Recherches sur son diagnostic et sur les méthodes de lutte. Convegno sulla Flavescenza dorata della vite, Vicenza-Verona, 28 maggio 1987, 175-208.

Carey J.R., (2003), Longevity: the biology and demography of life span, Princeton University Press, NJ.

Castañera M.B., Aparicio J.P., and Gürtler R.E., (2003), A stage-structured stochastic model of the population dynamics of Triatoma infestants, the main vector of Chagas disease, Ecol. Modeling 162, 33–53.

Chuche J., and Thiéry D., (2009), Cold winter temperatures condition the egg-hatching dynamics of a grape disease vector, Naturwissenschaften 96:827–834.

Cushing J.M., (1998), An introduction to structured population dynamics, Philadelphia, Society for Industrial and Applied Mathematics, pp. 193.

Decante D., and van Helden M., (2006), Population ecology of Empoasca vitis (Göthe) and Scaphoideus titanus (Ball) in Bordeaux vineyards: Influence of migration and landscape, Crop Protection 25:696–704.

Doerthy J., (2004), PEST Model-independent parameter estimation, 5th Edition.

Elliott P., Wakefield J., Best N., Briggs D., (2001), Spatial Epidemiology: Methods and Applications, Oxford University Press.

Ferriss R.S., Berger P.H., (1993), A stochastic simulation model of epidemics of arthropod-vectored plant viruses. Phytopathology 83, 1269-1278.

Galetto L., Roggia C., Marzachì C., Alma A., Bosco D., (2009), Acquisition of flavescence dorée phytoplasma by Scaphoideus titanus Ball from recovered and infected grapevines of Barbera and Nebbiolo cultivars. Proceedings 16th Meeting of the International Council for the Study of Virus and Virus-like Diseases of the Grapevine (ICVG). Dijon, France, 31 August – 4 September 2009. Le Progrès Agricole et Viticole, hors série, Spécial Congrès ICVG, pp. 120-121.

Gotta P., Morone C., (2001), Flavescenza dorata in Piemonte: gli interventi sul territorio. Informatore Agrario 57 (17), 89-90.

IRPCM Phytoplasma/Spiroplasma Working Team – Phytoplasma taxonomy group, (2004), International Journal of Systematic and Evolutionary Microbiology 54, 1243-1255.

Jeger M.J., (2000), Theory and plant epidemiology. Plant Pathology 49, 651-658.

Jeger M.J., Holt J., van Den Bosch F., Madden L.V., (2004), Epidemiology of insect-transmitted plant viruses: modelling disease dynamics and control interventions. Physiological Entomology 29, 291–304.

Levenberg K., (1944). A Method for the Solution of Certain Non-Linear Problems in Least Squares. The Quarterly of Applied Mathematics 2: 164–168.

Lessio F., Alma A., (2004), Dispersal patterns and chromatic response of Scaphoideus titanus Ball (Homoptera Cicadellidae), vector of the phytoplasma agent of grapevine flavescence dorée. Agricultural and Forest Entomology 6(2):121-127.
Lessio F., Tedeschi R., Alma A., (2007), Presence of Scaphoideus titanus on American grapevine in woodlands, and infection with "flavescence dorée" phytoplasmas. Bulletin of Insectology 60, 373–374.

Maanen van A., Xu X.-M., (2003), Modelling Plant Disease Epidemics. European Journal of Plant Pathology 109(7), 669-682.

Madden L.V., Jeger M.J., van den Bosch F., (2000), A theoretical assessment of the effects of vector-virus transmission mechanism on plant virus epidemics. Phytopathology 90, 576-594.

Madden L.V., van den Bosch F., (2002), A Population-Dynamics Approach to Assess the Threat of Plant Pathogens as Biological Weapons against Annual Crops. BioScience 52(1), 65-74.

Madden L.V., Hughes G., van den Bosch F., (2007), The study of plant disease epidemics, APS Press.

Maggi F., Marzachi' C., Bosco D., (2013), A Stage-Structured Model of Scaphoideus titanus in Vineyards. Environmental Entomology 42(2), 181-193.

Marzachì C., Boarino A., Vischi A., Palermo S., Morone C., Loria A., Boccardo G., (2001), Flavescenza dorata, legno nero e giallume dell'astro in vitigni del Piemonte sud orientale. Inf. Fitopatol. 9, 58-63.

Mood A., Graybill F.A., and Boes D.C., eds., (1974), Introduction to the theory of statistics, McGraw-Hill, pp. 564.

Metz J.A.J., and Diekmann O., (1986), The Dynamics of physiologically structured populations. Springer-Verlag, Lecture Notes in Biomathematics, pp. 511.

Morone C., Boveri M., Giosue S., Gotta P., Rossi V., Scapin I., Marzachi C., (2007), Epidemiology of Flavescence dorée in vineyards in northwestern Italy. Phytopathology 97, 1422-1427.

Murray J.D., (2002), Mathematical Biology; An Introduction, I., Interdisciplinary Appl. Math., 3rd. ed., Springer, New York.

Musetti R., Marabottini R., Badiani M., Martini M., Sanità di Toppi L., Borselli S., Borgo M., Osler R., (2007), On the role of H2O2 in the recovery of grapevine (Vitis vinifera cv. Prosecco) from Flavescence dorée disease. Functional Plant Biology 34, 750–758.

Nakazawa T., Yamanaka T. and Urano S., (2013), Model analysis for plant disease dynamics co-mediated by herbivory and herbivore-borne phytopathogens, Biol. Lett., doi:10.1098/rsbl.2012.0049

Ngwa G.A., (2006), On the Population Dynamics of the Malaria Vector, Bull. Math. Biol. 68:2161–2189.

Purcell A.H., (1982), Insect vector relationships with procaryotic plant pathogens. Annual Review of Phytopathology 20, 397-417.

Rigamonti I.E., Jermini M., Fuog D., and Baumgartner J., (2011), Towards an improved understanding of the dynamics of vineyard-infesting Scaphoideus titanus leafhopper populations for better timing of management activities , Pest Manag. 67:1222-1229.

Saltelli A., Ratto M., Andres T., Campolongo F., Cariboni J., Gatelli D., Saisana M., Tarantola S., (2008), Global Sensitivity Analysis, The Primer, John Wiley & Sons Ltd.

Schvester D., Carle P., Moutous G., (1963), Transmission de la flavescence dorée de la vigne par Scaphoideus littoralis Ball. Annales des Epiphyties 14, 175–98.

Sisterson M.S., (2009), Transmission of Insect-vectored Pathogens: Effect of vector Fitness as a Function of Infectivity Status. Environmental Entomology 38(2), 345-355.

Shtienberg D., (2000), Modelling: the basis for rational disease management. Crop Protection 19, 747-752.
Steffek R., Reisenzein H., Strauss G., et al., (2011), VitisCLIM, a project modeling epidemiology and economic impact of grapevine 'flavescence dorée phytoplasma' in Austrian viticulture under a climate change scenario. Bulletin of Insectology 64, S191-S192.

Truscott J.E., Webb C.R., Gilligan C.A., (1997), Asymptotic analysis of an epidemic model with primary and secondary infection. Bulletin of Mathematical Biology 59, 1101-1123.

Vandermeer J., Power A., (1990), An epidemiological model of the corn stunt system in Central America. Ecological Modelling 52, 235-248.

van der Plank J.E., (1960), Analysis of epidemics, In Horsfall J. Gand Cowling E.B. (eds), Academic Press, New York, USA, Plant Pathology, An Advance Treatise 3, 229–289.

van der Plank J.E., (1965), Dynamics of epidemics of plant disease. Science 147, 120-4.
Vandermeer J. and Power A., (1990), An epidemiological model of the corn stunt system in Central America. Ecological Modelling 52, 235-248.

Vidano C., (1964), Scoperta in Italia dello Scaphoideus littoralis Ball cicalina americana collegata alla "Flavescence dorée" della Vite. L'Italia agricola 101, 1031-1049.

Vitali M., Chitarra W., Galetto L., Bosco D., Marzachì C., Gullino M.L., Spanna F., Lovisolo C., (2013), Flavescence dorée phytoplasma deregulates stomatal control of photosyntesis in Vitis vinifera. Annals of Applied Biology 162, 335–346.

Weintraub P.G., Beanland L.-A., (2006), Insect vectors of phytoplasmas. Annu. Rev. Entomol. 51, 91–111.